\documentclass[final,5p,times,twocolumn]{elsarticle}

\usepackage{lineno,hyperref}
\usepackage{upgreek}
\usepackage[parfill]{parskip}
\usepackage[margin=5pt,font=small,labelfont=bf]{caption}
\usepackage{natbib}
\usepackage[none]{hyphenat}
\usepackage{amsthm}
\usepackage{amsmath}
\usepackage{array}
\usepackage{color}
\usepackage{multirow}
\usepackage{graphics}
\usepackage{wrapfig}
\usepackage{float}
\usepackage{placeins}
\usepackage{rotating,booktabs}
\usepackage{multirow}
\usepackage{lipsum}

\journal{Icarus}

\biboptions{authoryear}

\begin{document}

\begin{frontmatter}

%\title{New Results From Observations of the Comet D/Shoemaker-Levy 9 and the 2009 Wesley Impacts on Jupiter: Constraining the Impactor Origins}

\title{A retrospective analysis of mid-infrared observations of the Comet D/Shoemaker-Levy 9 and Wesley impacts on Jupiter}
\author[jpl]{J. A. Sinclair}\corref{cor1}
\ead{james.sinclair@jpl.nasa.gov}
\author[apl]{C. M. Lisse}
\author[jpl]{G. S. Orton}
\author[mich]{M. Krishnamoorthy}
\author[Leic]{L. N. Fletcher}
\author[harv]{J. Hora}
\author[FIT]{C. Palotai}
\author[Gem]{T. Hayward}

\address[jpl]{Jet Propulsion Laboratory/California Institute of Technology, 48Oak Grove Dr, Pasadena, CA 91109, United States}
\address[apl]{Applied Physics Laboratory, Johns Hopkins University, Baltimore, MD, United States}
\address[mich]{University of Michigan, Ann Arbor, MI, United States}
\address[Leic]{School of Physics \& Astronomy, University of Leicester, University Road, Leicester, LE1 7RH, United Kingdom}
\address[harv]{Center for Astrophysics, Harvard \& Smithsonian, Cambridge, MA, United States}
\address[FIT]{Florida Institute of Technology, Melbourne, FL, United States}
\address[Gem]{Gemini Observatory, NOIRLab, La Serena, Chile }

\begin{abstract}

We present a retrospective analysis of Earth-based mid-infrared observations of Jupiter capturing the aftermath of the impacts by Comet D/Shoemaker-Levy 9 (henceforth SL9) in July 1994 and the unknown object previously termed the “Wesley impactor” in July 2009.  While the effects of both impacts on Jupiter's atmosphere have been reported previously, we were motivated to re-examine both events using consistent data reduction and analysis methods to enable robust, quantitative comparisons.  To study the aftermath of the SL9 impacts, we examined infrared (7.8 - 20.5 $\upmu$m) spectrophotometry of Jupiter measured by the MIRAC (Mid-Infrared Camera) on NASA’s Infrared Telescope Facility on July 20, 21 1994 and low-resolution (R = 100) N-band (7 – 13 $\upmu$m) spectroscopy measured by SpectroCam-10 on the Palomar Telescope on July 22 1994.  To study the aftermath of the Wesley impact, we examined low-resolution (R = 100) N- and Q-band (R = 80, 17 – 25 $\upmu$m) spectroscopy recorded by Gemini South Telescope’s T-ReCS (Thermal-Region Camera Spectrograph). We analyzed the observations with two independent analyses: 1) a least-squares search over a grid of candidate mineral species to determine the composition of impact residue and 2) a radiative transfer analysis to derive atmospheric information.  We observe that the SL9 impact sites are enhanced in stratospheric CH$_4$ emissions at 7.9 $\upmu$m, due to shock heating and adiabatic compression from plume re-entry, and from 8.5 - 11.5 $\upmu$m due to stratospheric NH$_3$ emission and non-gaseous cometary material, in agreement with previous work.  In the G impact site, we derive NH$_3$ concentrations of $5.7^{+4.5}_{-2.8}$ ppmv at 30 mbar.  In new findings, we find that the SL9 impact sites also exhibit a non-gaseous emission feature at 18 - 19 $\upmu$m.    The non-gaseous emission at 8.5 - 11.5 $\upmu$m and 18 - 19 $\upmu$m emission result is best reproduced by predominantly amorphous olivine and (obsidian) silica at similar abundances.  The Wesley impact site exhibits enhanced emissions from 8.8 - 11.5 $\upmu$m and 18 - 19 $\upmu$m.  We found this could be reproduced by predominantly amorphous olivine and stratospheric gaseous NH$_3$ at concentrations of 150$^{+338}_{-121}$ ppbv retrieved at 30 mbar. Stratospheric abundances of NH$_3$ are a factor of $\sim$40 higher in the SL9 impacts compared to the Wesley impact, which confirms the former reached deeper, NH$_3$-richer altitudes of Jupiter's atmosphere compared to the latter. The absence of silicas in the Wesley impact would place an upper limit of 10 km/s on the incident angle and $\sim$9$^\circ$ on the entry angle of the impactor such that shock heating associated with the impact did not reach temperatures required for silicates to be converted.  

\end{abstract}


\end{frontmatter}

%\linenumbers
\section{Introduction}

Jupiter has the second largest gravitational sink in our solar system (after the Sun) and therefore experiences a relatively high impact rate compared to the other planets.  In recent decades, several impacts have been detected in Jupiter.  Most were bolide flashes that left no detectable residual influence on Jupiter including the September 10, 2012 impact \citep{hueso_2013} and the April 2020 impact \citep{giles_2021} detected by Juno’s ultraviolet instrument (UVS, \citealt{gladstone_uvs_2017}).  However, two impact events left prominent perturbations to Jupiter’s atmosphere. One was the series of impacts by fragments of Comet D/Shoemaker-Levy 9 (hereafter SL9) in July 1994 (c.f. the review by \citealt{harrington_2004}).  The other was an impact by an unknown body into the nightside of Jupiter on July 19, 2009, the aftermath of which was first detected by amateur astronomer Anthony Wesley (hereafter the Wesley impact, c.f. \citealt{sanchez_lavega_2010}).  Using spectroscopic and imaging observations from Earth-based telescopes, the effects of both impact events on the atmosphere of Jupiter were investigated intensely.  

SL9 was tidally disrupted into at least 16 fragments, which then impacted Jupiter's atmosphere at $\sim$40$^\circ$S between July 16 and 22, 1994.  The impacts occured at relative velocities similar to Jupiter's escape velocity ($\sim$65 km/sec), shock-heated gas to extreme temperatures (30000 - 40000 K, e.g. \citealt{zahnle_1994}) and vaporized the cometary material.  The residue then rose in a plume along the initial entry path, outside of the atmosphere, before re-entering the atmosphere at near-horizontal trajectories (e.g. \citet{griffith_1997}, Figure 14).  Upon re-entry, the material re-compressed and began sinking in atmosphere, converting kinetic to thermal energy and thereby heating the atmosphere as deep as the lower stratosphere and enhancing hydrocarbon emissions at mid-infrared wavelengths (e.g. \citealt{orton_1995,bezard_1997}).  Species such as NH$_3$, normally cold-trapped below Jupiter's tropopause,  were excavated to stratospheric altitudes and observed in emission at mid-infrared wavelengths (e.g. \citealt{kostiuk_1996,griffith_1997}).   Shock-induced chemistry and/or delivery from the comet itself enhanced the abundances of existing species, such as HCN  (e.g. \citealt{bezard_1997}, \citealt{noll_1995}) and introduced new chemical species into the stratosphere including H$_2$O (e.g. \citealt{bjoraker_1996}, \citealt{encrenaz_1997}) and CO (e.g. \citealt{lellouch_1997}, \citealt{kim_1999}).  In particular, stratospheric H$_2$O and CO continue to remain detectable in Jupiter's stratosphere at the time of writing (e.g. \citealt{lellouch_2002}, \citealt{cavalie_2013}, \citealt{benmahi_2020}). 

The aftermath of the Wesley impact was first noticed on July 19 2009 at $\sim$55$^\circ$S as a dark ``bruise'' at visible wavelengths  \citep{sanchez_lavega_2010} but also bright at near-infrared wavelengths \citep{hammel_2010,dePater_2010,orton_2011} indicating the presence of material at millibar pressure levels.  The same region exhibited stratospheric NH$_3$ emission at mid-infrared wavelengths (e.g. \citealt{fletcher_impact_2011}, \citealt{dePater_2010}), which indicated an impactor had excavated NH$_3$ from below the tropopause to stratospheric altitudes, similar to the SL9 impacts.  By dynamical arguments, impacts on Jupiter are orders of magnitude more likely to be comets than asteroids since Jupiter should have cleared the latter from its orbit $\sim$Gyr ago \citep{schenk_2004}.  Nevertheless, there was a convergence of evidence suggesting the Wesley impactor was an asteroid.  Several studies inferred the material from the Wesley impactor was more ``rocky'' and less icy than material from the previous SL9 impacts.  Using ultraviolet observations measured by the Hubble Space Telescope (HST), \citet{hammel_2010} noted that the 2009 debris field was less diffuse and shrank faster than comparable-sized SL9 fields, which implied the Wesley impactor material was heavier and denser and therefore sank faster in the atmosphere than the SL9 material.  While strong, stratospheric heating was observed for several days following the SL9 impacts (see above), no such heating occured in the aftermath of the Wesley impact, as evidenced by the lack of any enhanced, stratospheric CH$_4$ emissions (e.g. \citealt{fletcher_impact_2010}, \citealt{orton_2011}).  This was interpreted as the impactor material being sufficiently heavier/denser such that the plume did not rise as high as the upper stratosphere.  Enhanced, stratospheric C$_2$H$_6$ emissions were observed \citep{fletcher_impact_2010}, which indicated an increase in its abundance within the impact region.  An enhancement of C$_2$H$_6$, in contrast to oxidized products such as H$_2$O and CO, was more consistent with a ``dry'' impactor with a higher C/O ratio \citep{zahnle_1996}, as expected for an asteroid.  From the discovery images of the 2009 impact, \citet{sanchez_lavega_2010} inferred an impact trajectory elevation angle of 20$^\circ$ $\pm$ 5$^\circ$, much shallower than the 45$^\circ$ angle of SL9. Assuming ballistic trajectories, similar to the approach of \citet{pankine_1999}, and comparing the visible debris field with the ``intermediate" SL9 fragments, they estimated the impactor size as 0.5-1 km.  These measurements left few constraints on the trajectory of the impactor, but \citet{orton_2011} examined a suite of possible trajectories and determined that (a) an asteroidal origin for the impactor was possible, and (b) of the most likely sources of asteroids ejected into Jupiter-encountering orbits - Trojans (\citealt{levison_1997}, \citet{duncan_2004}) or Main-Belt Hildas \citep{disisto_2005} - that the Hilda family was more likely.  

A further contrast between the two impact events was the detection of a 9.1-$\upmu$m non-gaseous feature in the Wesley impact region, attributed to crystalline silicas such as quartz and cristobalite \citep{orton_2011, fletcher_impact_2011} and its (apparent) absence from the SL9 impact.  Stringent upper limits of silica of 0.5\% in abundance have been derived from high signal-to-noise ratio spectra of comets, including comets 9P/Tempel 1, C/Hale-Bopp 1995 O1 \citep{lisse_2007a} and 17P/Holmes \citep{reach_2010}.   Similarly, there is no spectroscopic evidence of silica on asteroids or Trojans (e.g. \citealt{emery_2006}, \citealt{vernazza_2010}).  The presence of silicas in the residue of the Wesley object was therefore interpreted to have been formed during the impact by the high-pressure ($>$10$^4$ bar) and/or high-temperature ($>$ 1500 K) alteration of ferromagnesian silicates, as seen in terrestrial studies of tektites, silica magmas and rock fulgurites \citep{lisse_2009}.  However, such temperatures and pressures were achieved during the SL9 impacts (e.g. \citealt{maclow_1994}, \citealt{deming_2001},  \citealt{korycansky_2006}).  Amorphous olivine and amorphous pyroxene (typically found in equal abundances in comets) would have been readily converted to silicas and so the apparent absence of silicas from the SL9 impacts was puzzling.

Overall, contrasts between the SL9 and Wesley impacts led to the inference that the Wesley impactor must be different in composition and/or origin from SL9.  However, previous observations of the SL9 and Wesley impacts employed different methods for data reduction, calibration and radiative transfer calculations and so previous comparisons between the events were not immediately robust.  We were therefore motivated to re-examine and quantitatively compare both impacts using consistent methods.  In this work, we perform an analysis of Earth-based, mid-infrared observations recorded following the impacts of SL9's fragments in July 1994 and the Wesley impact in July 2009.

\section{Observations}\label{sec:obs}
\subsection{SL9 impact observations in 1994}
\subsubsection{IRTF-MIRAC}

\begin{figure*}[h]
\begin{center}
\includegraphics[width=0.8\textwidth]{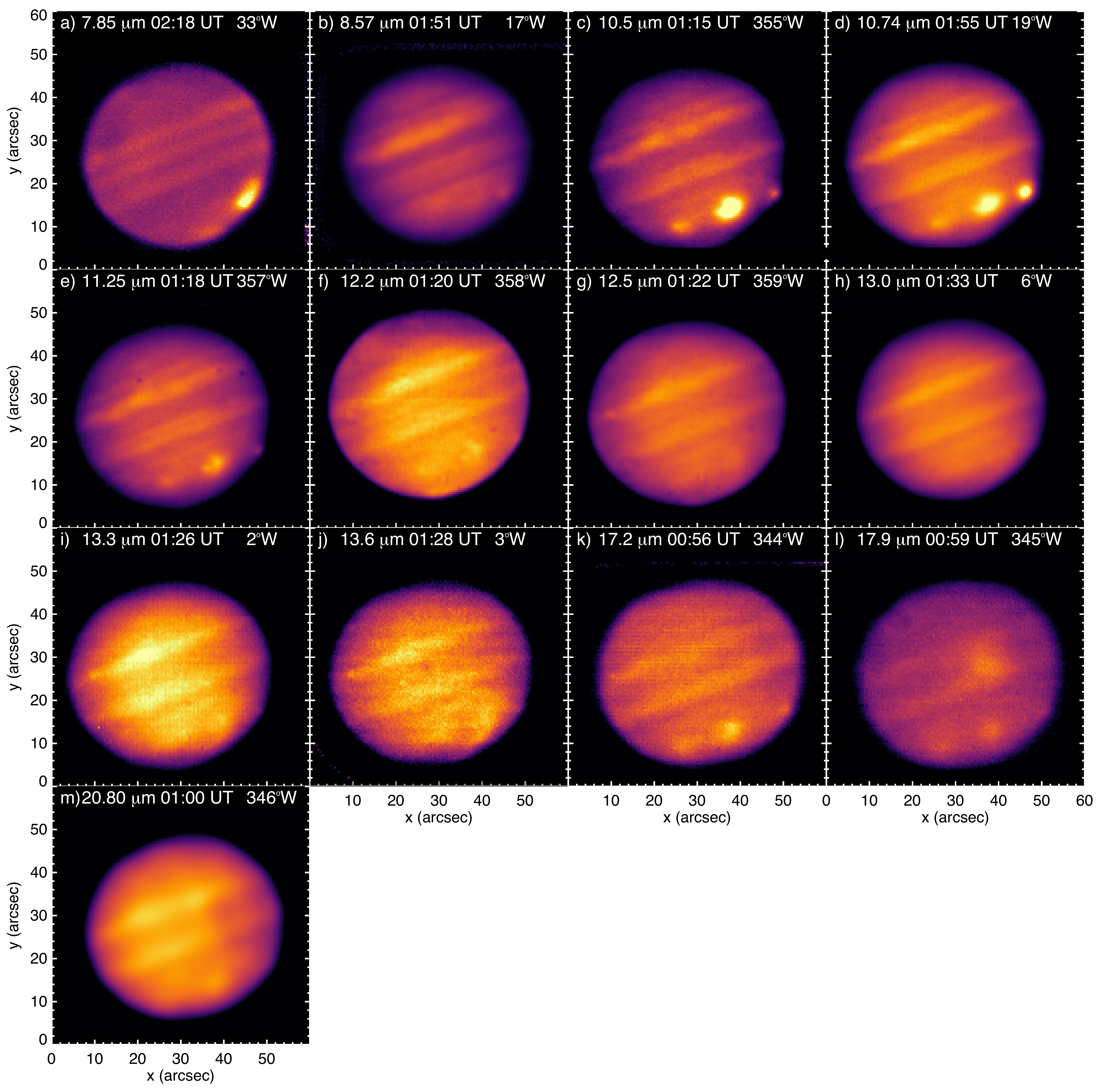}
\caption{IRTF-MIRAC images recorded on 1994 July 20.  The filter wavelength, time of observation (UT) and central meridian longitude (System III) are included in each panel.  The aftermath of SL9 impacts are evident as bright regions on the disk, particularly at 7.85, 10.50 – 11.25 and 17.20 $\upmu$m.   }
\label{fig:mirac_jul20}
\end{center}
\end{figure*}
\begin{figure*}[t]
\begin{center}
\includegraphics[width=0.85\textwidth]{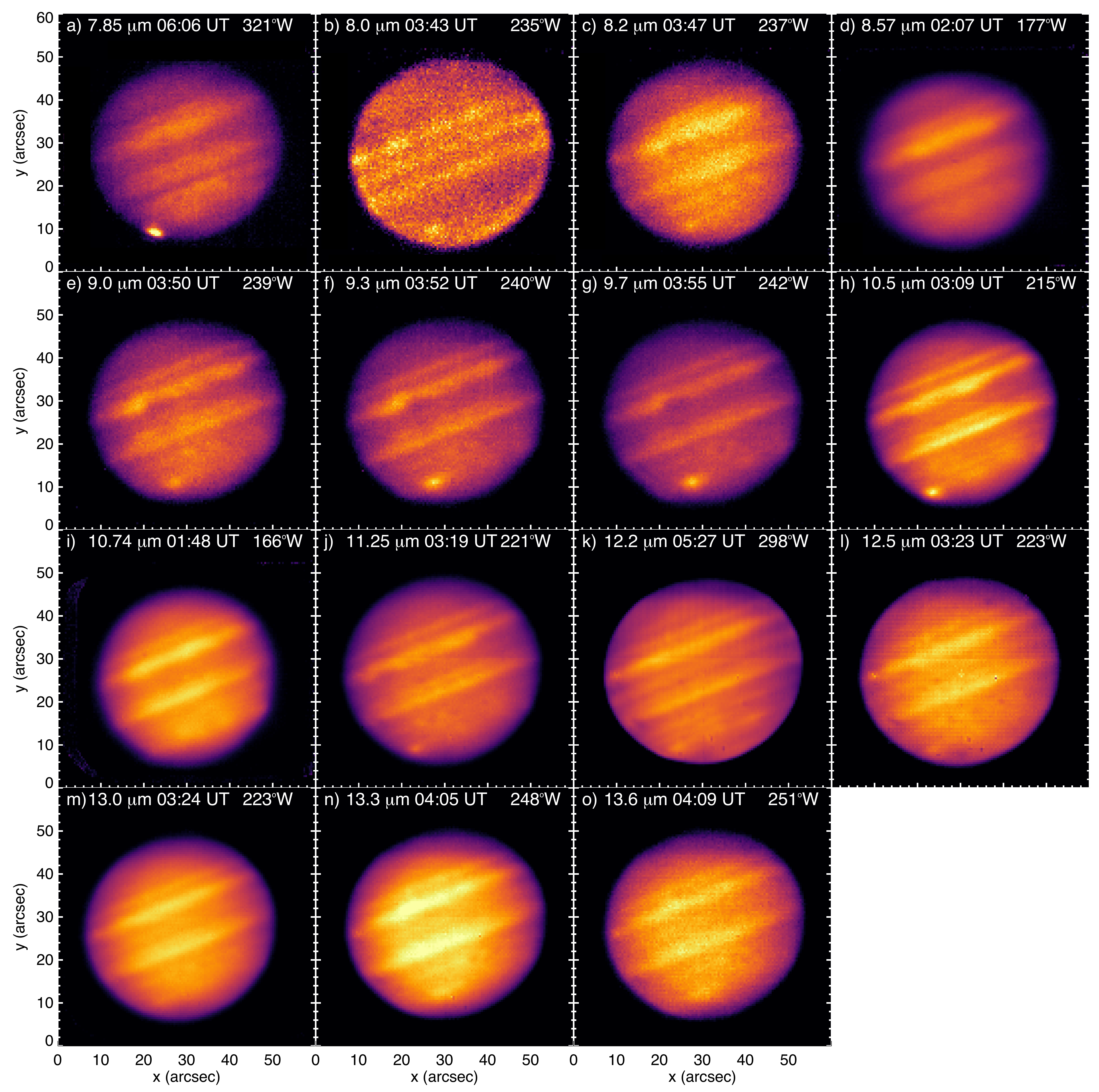}
\caption{As in Figure \ref{fig:mirac_jul20} but using images recorded on 1994 July 21.   }
\label{fig:mirac_jul21}
\end{center}
\end{figure*}

\begin{figure}[h!]
\begin{center}
\includegraphics[width=0.45\textwidth]{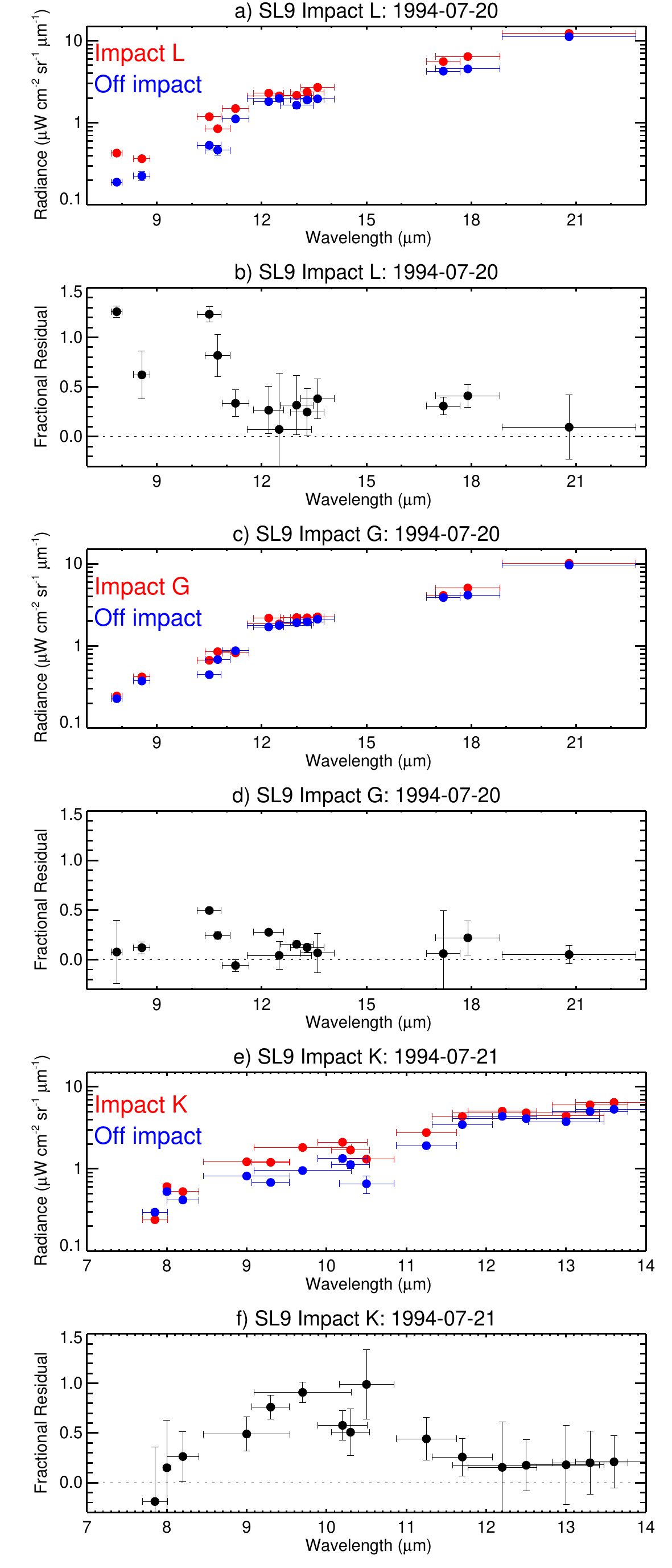}
\caption{a) IRTF-MIRAC spectrophotometry of the SL9 impact site L (red points) and background atmosphere (blue points) on 1994 July 20, b) the fractional residual between the impact site and background (Equation \ref{eq:frac_resid}).  Similar results are shown for SL9 impact G on the same date (c, d) and SL9 impact K on 1994 July 21 (e, f).  Horizontal error bars represent the full width at half maximum (FWHM) of the given filter. No Q band images were recorded on 1994 July 21.   }
\label{fig:fr_mirac}
\end{center}
\end{figure}
Spectrophotometric images from 7.85 to 20.5 $\upmu$m were measured using the Mid-Infrared Array Camera, MIRAC \citep{hoffmann_1998} on NASA’s 3-m Infrared Telescope Facility (IRTF). The data were reduced, using standard object-minus-sky (A-B) subtraction and flat-fielding steps.  Absolute calibration was performed by scaling the images to convolved, mid-infrared spectra recorded by the Voyager IRIS and Cassini CIRS instruments, following the approach used by \citet{fletcher_2009}. The noise-equivalent radiances (NESR) of the images were calculated by determining the standard deviation of sky pixels more than 5'' outside of Jupiter’s limb. 

We concentrated on observations of the impact sites of the G, K and L fragments, as they were bright and had the best wavelength coverage. Images recorded on 1994 July 20 captured the G and L impacts approximately 41 - 43 hours and 2 - 4 hours after their predicted times of impact, respectively. Images on July 21 captured the K impact 18 - 20 hours after the predicted time of impact. Figures \ref{fig:mirac_jul20} and \ref{fig:mirac_jul21} show the images recorded on July 20 and 21 and Table \ref{tab:sl9_obs} further details each image and the time elapsed since the G, K and L impacts.

\begin{table*}[!ht]
\footnotesize
\centering
\begin{tabular}{>{\centering\arraybackslash} m{1.7cm} |>{\centering\arraybackslash} m{2.4cm}  | >{\centering\arraybackslash} m{1.1cm}  >{\centering\arraybackslash} m{1.2cm} >{\centering\arraybackslash} m{1.0cm} >{\centering\arraybackslash} m{1.5cm}  >{\centering\arraybackslash} m{1.2cm} >{\centering\arraybackslash} m{1.2cm} >{\centering\arraybackslash} m{1.2cm} }
Date & Instrument & Time & $\lambda$ & Airmass & Longitude & $\Delta t_G $ & $\Delta t_K $ & $\Delta t_L $ \\
\multirow{13}{*}{1994-Jul-20} & \multirow{13}{*}{IRTF/MIRAC} &   (UT)      &   ($\upmu$m)  &   & range &  (hrs)     & (hrs) & (hrs)\\
\firsthline 
&  & 00:56 & 17.2 &       2.34 &       264 - 64 &       41.34 & - &       2.57 \\
&  & 00:59 & 17.9 &       2.29 &       265 - 65 &       41.39 & - &       2.62 \\
&  & 01:00 & 20.80 &       2.27 &       266 - 66 &       41.41 & - &       2.63 \\
&    & 01:15 & 10.5 &       2.05 &       275 - 75 &       41.66 & - &       2.88 \\
&      & 01:18 & 11.25 &       2. &       277 - 77 &       41.71 & - &       2.93 \\
&  & 01:20 & 12.2 &       1.98 &       278 - 78 &       41.74 & - &       2.97 \\
&  & 01:22 & 12.5 &       1.96 &       279 - 79 &       41.78 & - &       3.\\
&    & 01:26 & 13.3 &       1.90 &      78 - 278 &       41.84 & - &       3.07 \\
&  & 01:28 & 13.6 &       1.89 &      77 - 277 &       41.88 & - &       3.10 \\
&  & 01:33 & 13.0 &       2.15 &      74 - 274 &       41.96 & - &       3.18 \\
&  & 01:51 & 8.57 &       2.11 &      63 - 263 &       42.26 & - &       3.48 \\
&  & 01:55 & 10.74 &       2.02 &      61 - 261 &       42.33 & - &       3.55 \\
&  & 02:18 & 7.85 &       2.19 &      47 - 247 &       42.71 & - &       3.93 \\
\hline 

\multirow{18}{*}{1994-Jul-21}  & \multirow{18}{*}{IRTF/MIRAC}      & 01:48 & 10.74 &       2.44&       86 - 246 & - &     15.28& -  \\
&& 02:07 & 8.57 &       2.26&       97 - 257& -  &     15.6& -  \\
&  & 03:09 & 10.50 &       1.30&       135 - 295 & -  &    16.63& -  \\
&    & 03:19 & 11.25 &       1.26&       141 - 301 & -  &    16.8& -  \\
&      & 03:23 & 12.50 &       1.25&       143 - 303 & -  &    16.86& -  \\
&        & 03:24 & 13.&       1.57&       143 - 303 & -  &    16.88& -  \\
&  & 03:43 & 8.00&       1.22&       155 - 315 & - &     17.2& -  \\
&  & 03:47 & 8.20 &       1.21&       157 - 317 & - &     17.26& -  \\
&  & 03:50 & 9.00&       1.21&       159 - 319 &  - &    17.31& -  \\
&  & 03:52 & 9.30 &       1.20&       160 - 320 & - &     17.35& -  \\
&  & 03:55 & 9.70 &       1.20&       162 - 322 & -  &    17.4& -  \\
&  & 03:58 & 10.20 &       1.20&       164 - 324 &  - &    17.45& -  \\
&    & 04:01 & 11.70 &       1.20&       166 - 326 & - &     17.5& -  \\
&  & 04:05 & 13.30 &       1.19&       168 - 328 & -  &    17.56& -  \\
&  & 04:09 & 13.60 &       1.19&       171 - 331 &  - &    17.63& -  \\
&    & 05:27 & 12.20 &       1.19&       218 - 378 & - &     18.93& -  \\
&  & 05:31 & 10.30 &       1.20&       220 - 380 &  -  &   19.0& -  \\
&& 06:06 & 7.85 &       1.28&       241 - 401 &   -   & 19.58& -  \\

\hline
1994-Jul-22 & Palomar/SC-10 & 01:49 & 8 - 13.5 & 1.64 & 308 - 35 & 90.23 & - & -  \\ 
\lasthline
\end{tabular}
\caption{Details of the observations examining the SL9 G, K and L impact sites.   $\Delta t_G $,  $\Delta t_K $ and $\Delta t_L $ denote the time elapsed (in hours) since impacts G, K and L.  }
\label{tab:sl9_obs}
\end{table*}

From each image recorded on July 20, a broadband spectrum of the L impact was derived by calculating the mean radiance and emission angle between 35-45$^\circ$S (planetocentric) and 330-343$^\circ$W (System III).  The uncertainty on the mean radiance was calculated as the larger of: 1) the standard deviation on the mean or 2) the NESR scaled by n$_{pixel}^{0.5}$, where n$_{pixel}$ is the number of diffraction-resolved pixels averaged.  A similar broadband spectrum of the G impact site was extracted from the July 20 images at same latitude range but longitudes from 23 – 32$^\circ$W.  Using the images recorded on July 21, a broadband spectrum of the K impact was similarly extracted using longitudes from 260-275$^\circ$W in the same latitude range. 

We considered several methods for calculating the radiance of the unperturbed “background” atmosphere from each image in order to quantify the enhancement in emission due to the impacts.  Ultimately, we chose to derive this background ``off-impact'' radiance by calculating the mean radiance within the same longitude range of the impacts (see above) but over a latitude range of 28-38$^\circ$N.  The chosen latitude range was determined such that the radiances of the impact site and the “background” unperturbed site were as similar as possible in emission angle, such that both observations sound a similar altitude level in Jupiter’s atmosphere and such that foreshortening effects are removed when computing a ratio between the two.  We chose not to calculate the background atmosphere using unperturbed areas within the same 35-45$^\circ$S latitude band.  By July 20 and July 21, over 9 individual fragments of SL9 had impacted the atmosphere and it was near-impossible to extract radiances of an unperturbed region at a similar emission angle as the impact site. While MIRAC images were recorded on July 14 before the SL9 impacts, we chose not to use these images to calculate the background atmosphere for the following reasons.  Firstly, the images on July 14, July 20 and 21 captured different longitude ranges of Jupiter and thus the sites of the future impacts were either not sampled on July 14 or recorded at very different emission angles compared to the July 20 - 21 images.  Secondly, images on July 19/20 were recorded in several CVF (circular-variable filter) positions between 8 and 12 $\upmu$m, which were not recorded on July 14 images.  Limiting the study of the K and L impact sites in July 19-20 images only to the filters used on July 14 would remove measurements at key wavelengths where non-gaseous species have spectral features.  

In order to quantify the enhancement in emission due to impact, we calculated the fractional residual, $f_r$, between radiances measured over the impact region, $R_i$, and the off-impact/background region, $R_b$, as defined in Equation \ref{eq:frac_resid}.

\begin{equation}\label{eq:frac_resid}
f_r = \frac{R_i - R_b}{R_b}
\end{equation}

Figure \ref{fig:fr_mirac} shows the broadband spectra of the G, L and K impact sites, the background atmosphere as detailed above, and the fractional residual between them (Equation \ref{eq:frac_resid}).   As noted in previous work, the site of the L impact is enhanced at 7.85 $\upmu$m due to CH$_4$ emissions enhanced by stratospheric fallback heating.  Marginal enhancements between 12 and 13 $\upmu$m are noted and are attributed to emissions of C$_2$H$_6$ and C$_2$H$_2$, which are also enhanced by stratospheric heating and possibly by enhancements to their abundances.   In both the K and L impacts, the 8.5 – 11.5 $\upmu$m range is enhanced due, in part, to emissions from gaseous NH$_3$ lofted into the stratosphere as well as non-gaseous compounds produced from the cometary debris upon impact. Significant enhancements in emission are also observed at ~18 - 19 $\upmu$m for the G and L impacts, whereas impact and background radiances at 20.5 $\upmu$m agree within uncertainty.  As far as we are aware, there are no gases in Jupiter’s atmosphere that have significant spectral features at this wavelength and so we attribute this feature to non-gaseous compounds produced from the cometary debris.  For the G and K impact, radiances at 7.85 $\upmu$m and 12 – 14 $\upmu$m agree with background radiances within uncertainty, which indicates less stratospheric heating or chemical alteration relative to the L impact site.  This suggests that the G and K fragments were smaller than the L impactor fragment and therefore produced a smaller atmospheric response. Alternatively, the G and K impact sites had evolved and cooled in the 18-20 and 41-43 hours between the predicted times of their impacts and their respective measurements.  In contrast, the site of the L impact was observed only 2 – 4 hours after the predicted impact time.  

\subsubsection{Palomar/SpectroCam-10}

\begin{figure}[t]
\begin{center}
\includegraphics[width=0.45\textwidth]{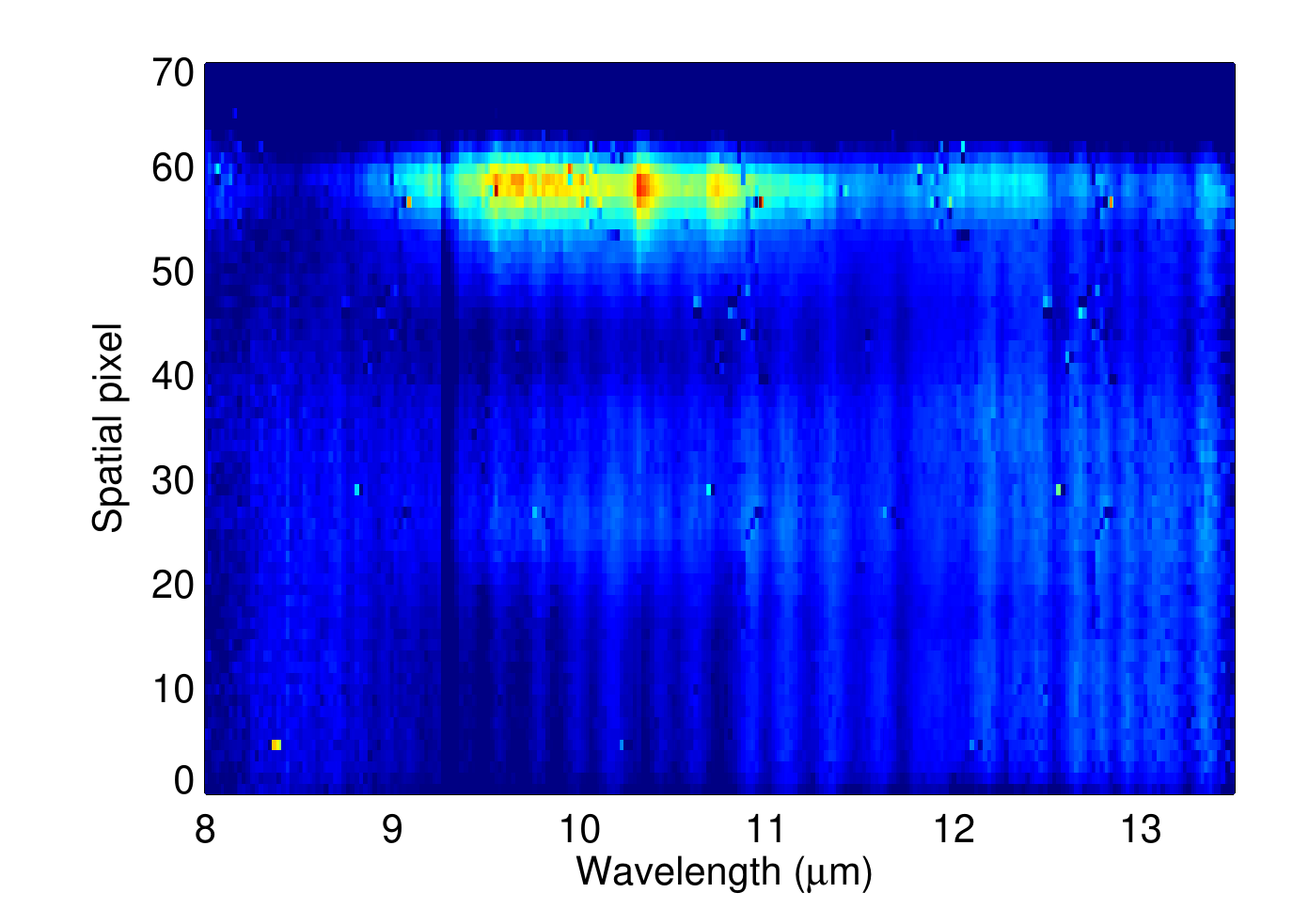}
\caption{The spectral-spatial image of the Palomar/SC-10 observation recorded on 1994 July 22.  Spatial pixels 0 – 50 capture the unperturbed atmosphere west-to-east in the $\sim$44$^\circ$S latitude band , the enhanced emissions of SL9 impact G are evident near the limb at spatial pixels ~55-60 and dark sky off planet is captured from spatial pixels 62 – 70. }
\label{fig:palomar_G}
\end{center}
\end{figure}

We also examined spectroscopic measurements recorded on 1994 July 22 by the SpectroCam-10 (henceforth `SC-10') instrument \citep{mcghee_2000} at the 200-inch (5.1 m) Hale telescope at the Palomar Observatory.  These spectra capture the site of the SL9 G impact approximately 90 hours  after the predicted time of impact.   Table \ref{tab:sl9_obs} provides further details of the Palomar/SC-10 observations recorded.

We did not examine spectra from the R-impact site, as they were largely dominated by blackbody emission with fewer identifiable spectral features and have already been presented by \citet{nicholson_1995}.  Spectra were obtained using the 1x15'' slit with the slit length oriented parallel to Jupiter’s equator and covering the region of peak brightness due to the impact.  The slit position also captured 5 – 6 sky pixels off Jupiter from which the standard deviation was calculated to derive the noise-equivalent radiance.  Each exposure was reduced by subtracting an off-Jupiter exposure. Bad pixels in the image were removed by interpolation. A slight tilt of the spectral lines in the images was corrected such that they were aligned vertically in each image. The wavelength grid was derived using the known wavelength of telluric ozone.  Figure \ref{fig:palomar_G} shows the resulting spectral-spatial image.  

Spatial registration and viewing geometries of each pixel were calculated by identifying the location of Jupiter’s limb in the spectral-spatial image (Figure \ref{fig:palomar_G}), the known latitude of the G impact (44$^\circ$S), the 0.25'' pixel scale of the instrument and the sub-observer latitude and longitude at the time of observation determined by JPL Horizons. 

Initially, the absolute calibration was performed using Callisto, just as for the R-impact observations reported by \citet{nicholson_1995}. However, this resulted in radiances (even away from the impact) that were factors of 3 – 4 higher than typical values for Jupiter.  We suggest differences in atmospheric seeing between the measurements of Jupiter and Callisto could account for these unphysically-large radiances. In order to calibrate the Palomar/SC-10 data in absolute radiance, we instead scaled the observed radiances to CIRS (Composite Infrared Spectrometer, \citealt{kunde_1996}) spectra of Jupiter recorded during the 2000-2001 flyby.  Using the 2.5 cm$^{-1}$ resolution (0.025 $\upmu$m resolution at 10 $\upmu$m) CIRS data, we computed a mean spectrum between 40-45$^\circ$S over all sampled longitudes. The CIRS spectrum was then convolved down to the coarser 0.1 $\upmu$m (10 cm$^{-1}$) resolution of Palomar/SC-10. We coadded Palomar/SC-10 spectra over 10 spatial pixels centered over the central meridian, and away from the impact region, which resulted in a mean emission angle similar to the CIRS spectrum.  A calibration scale factor was derived by computing a ratio of the mean CIRS radiance to the mean Palomar count from 8.9 to 13 $\upmu$m, and subsequently applied to all Palomar radiances.  

\begin{figure}[t]
\begin{center}
\includegraphics[width=0.5\textwidth]{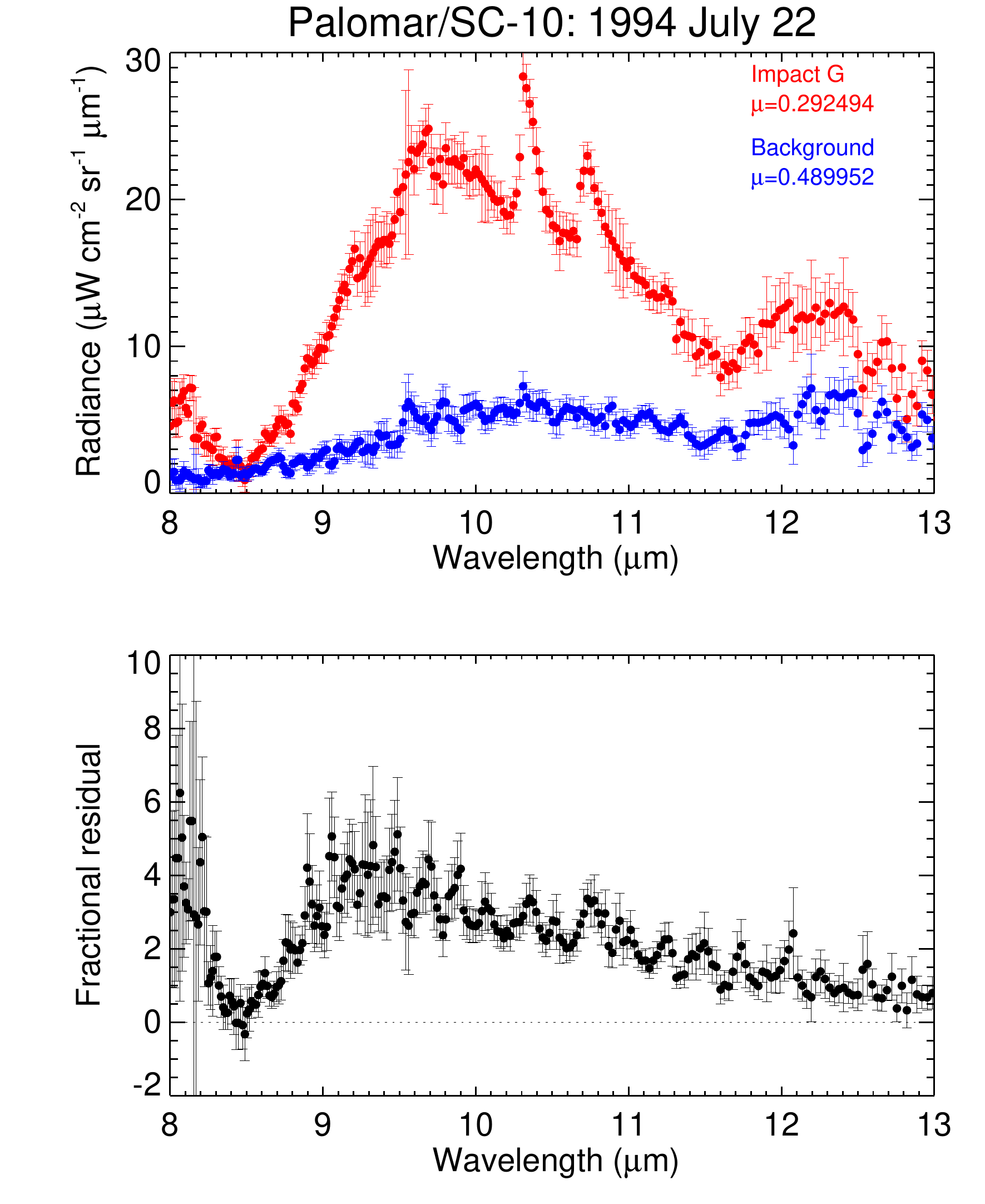}
\caption{The top panel shows Palomar/SpectroCam-10 spectra of the SL9 G impact site (red) and an unperturbed region in the same latitude band (blue).  The lower panel shows the fractional residual between the two spectra.  Note the difference in emission angles of the two observations. }
\label{fig:palomar_spx}
\end{center}
\end{figure}

A coadded spectrum of the G impact site was computed by averaging the spectra obtained in spatial pixels 53 – 56 (approximately longitudes of 23-32$^\circ$W), which resulted in a mean emission angle of 73$^\circ$ ($\upmu$ $\sim$ 0.3, where $\upmu$ is the cosine of the emission angle).  The 15'' slit of SC-10 samples dark sky off the eastern limb and the 44$^\circ$S latitude band east of the central meridian.  Unfortunately, no spectra were measured at $\upmu$ $\sim$ 0.3 west of the central meridian, which would allow a spectrum of the background atmosphere to be computed at a similar viewing geometry as the G impact site for comparison.  We coadded spectra over spatial pixels 40 – 50 (or longitudes of 0 – 18$^\circ$W) as a compromise between averaging a sufficient number of spectra to increase the effective signal-to-noise ratio and achieving a mean emission angle ($\upmu$ $\sim$ 0.5) as similar as possible to the impact spectrum.  We anticipate the difference in emission angle between the impact and non-impact region may introduce an offset in computing the fractional residual since different emission angles sound slightly different levels of the atmosphere (at different temperatures).  Nevertheless, we continued to analyze the spectra such that a spectroscopic analysis of both SL9 and Wesley impacts could be performed. 

Figure \ref{fig:palomar_spx} shows the Palomar spectra of the SL9 G impact site, background atmosphere and the fractional residual (Equation \ref{eq:frac_resid}) between the two locations.  As observed by MIRAC at other SL9 sites (Figure \ref{fig:fr_mirac}), the G impact site is enhanced in CH$_4$ emission at $\sim$8 due to stratospheric heating and between 9 – 11 $\upmu$m due to NH$_3$ gas lofted in to the stratosphere and non-gaseous compounds produced from the cometary material.  

\subsection{Wesley impact observations in 2009}

\begin{table*}[!t]
\centering
\begin{tabular}{>{\centering\arraybackslash} m{1.7cm} |>{\centering\arraybackslash} m{2.7cm}  | >{\centering\arraybackslash} m{1.1cm}  >{\centering\arraybackslash} m{1.2cm} >{\centering\arraybackslash} m{1.0cm} >{\centering\arraybackslash} m{1.5cm} }
Date & Instrument & Time & Spectrum & Airmass & CML \\
\firsthline 
\multirow{13}{*}{2009-Jul-24} & \multirow{13}{*}{Gemini-S/T-ReCS} & 04:25 & N1 & 1.176 & 278 \\
& &  04:31 & N2 & 1.160 & 281 \\
& &  04:38 & N3 & 1.145 & 285 \\
& &  04:45 & N4 & 1.132 & 290 \\
& &  04:51 & N5 & 1.119 & 293 \\
& &  04:57 & N6 & 1.108 & 297 \\
& &  05:05 & N7 & 1.097 & 302 \\
& & 06:35 & Q1 & 1.041 & 356 \\
& &  06:41 & Q2 & 1.042 & 360 \\
& &  06:48 & Q3 & 1.045 & 4 \\
& &  06:55 & Q4 & 1.048 & 8 \\
& &  07:01 & Q5 & 1.052 & 12 \\
& &  07:08 & Q6 & 1.057 & 16 \\
\lasthline
\end{tabular}
\caption{Details of the observations capturing the aftermath of the Wesley impact.  Spectra beginning with `N' and `Q' are N-band (7.5 - 13 $\upmu$m) and Q-band (17 - 25 $\upmu$m) spectra, respectively. `CML' denotes the central meridian longitude at the time of observation in System III. }
\label{tab:wesley_obs}
\end{table*}
\begin{figure*}[h!]
\begin{center}
\includegraphics[width=0.75\textwidth]{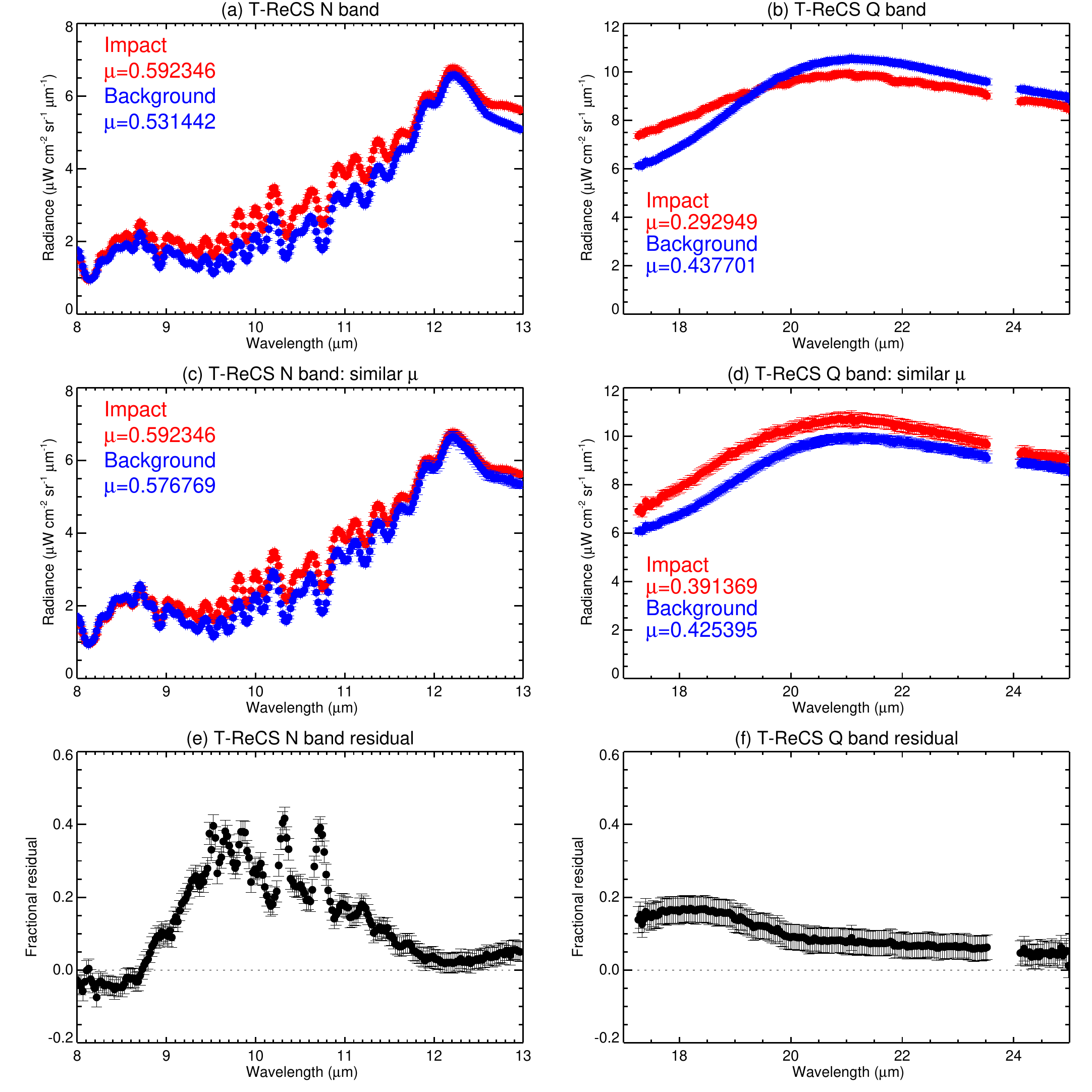}
\caption{T-ReCS N band (a) and Q band (b) coadded spectra of the Wesley impact (red spectra) and a background region for comparison (blue) as recorded on 2009 Jul 24.  The cosine of emission angle of the spectra ($\upmu$) are shown.  Panels (c) and (d) show similar spectra except where individual observations were omitted such that the coadded spectra of the impact and background were similar as possible in emission angle. Panels (e) and (f) show the fractional residual (Equation \ref{eq:frac_resid}) between the spectra shown in panels (c) and (d).   }
\label{fig:trecs_spx}
\end{center}
\end{figure*}
\begin{figure*}[ht]
\begin{center}
\includegraphics[width=0.8\textwidth]{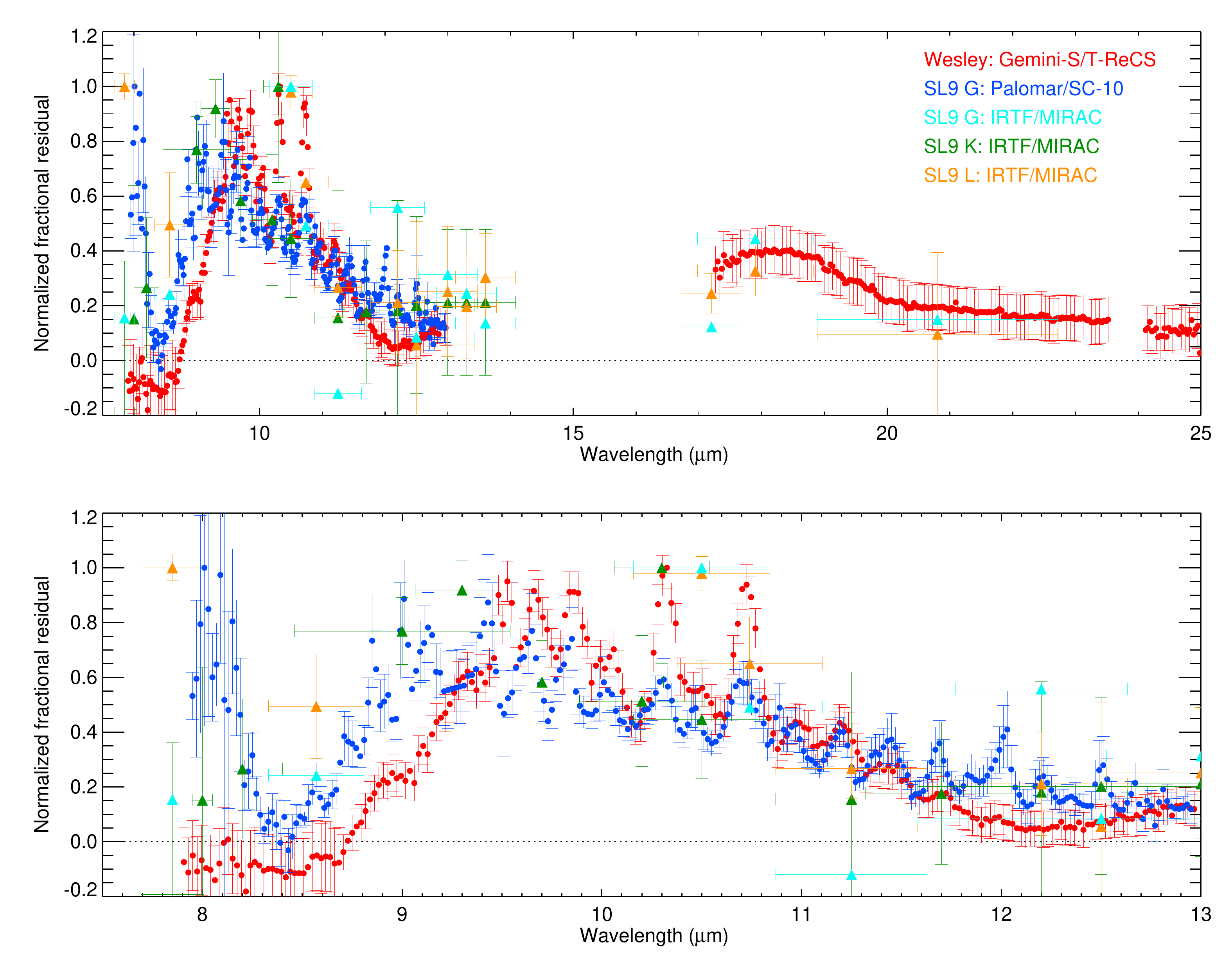}
\caption{Comparison of the fractional residuals of the SL9 and Wesley impact sites from all datasets.  The top panels shows the 7.5 - 25 $\upmu$m range, the bottom panel focuses on the 7.5 - 13 $\upmu$m range.  The fractional residuals have been normalized to unity for ease of comparison.   }
\label{fig:compare_all}
\end{center}
\end{figure*}
Low-resolution (R = 100) N-band (7.5-13 $\upmu$m) and (R = 80) Q-band (17 – 25 $\upmu$m) spectra were measured using the Thermal-Region Camera Spectrograph, T-ReCS, on the 8-m Gemini South Telescope on 2009 July 24, approximately four (Earth) days or $\sim$10 Jupiter rotations after the effects of the impact were first observed. The slit of the spectrograph was aligned parallel to Jupiter’s equator, centered over the latitude of the Wesley impact ($\sim$60$^\circ$S), and measurements were repeated while Jupiter rotated.  7 N-band and 6 Q-band measurements were recorded over the course of the night, providing multiple emission angle spectra of the impact site (with the core at $\sim$304$^\circ$W) as it rotated across Jupiter's disk as well spectra of the unperturbed atmosphere in the same latitude band.  The measurements are detailed in Table \ref{tab:wesley_obs}.  Spectra were also measured of the Cohen standard HD216032 \citep{cohen_1999}, which were used to perform the nominal, absolute calibration of the Jupiter spectra.  These spectra were previously analyzed and presented by \citet{fletcher_impact_2011} and \citet{orton_2011}.  Initially, they found that the N- and Q-band spectra could not be fit simultaneously with a sensible atmospheric model, which was attributed to the wavelength-dependent effects of telluric contamination.  In order to resolve this, \citet{fletcher_impact_2011} first calculated the ratio between spectra of the impact and background atmosphere such that telluric contamination is removed.  Second, they computed a series of synthetic spectra from 7 - 25 $\upmu$m over a range of emission angles using an atmospheric model derived from Cassini-CIRS (Composite Infrared Spectrometer, \citealt{kunde_1996}) measurements during the 2001 flyby \citep{fletcher_ph3_2009}.   The impact-to-background ratios were then applied to synthetic spectra (at the appropriate emission angle) to compute absolute spectra in radiance units without telluric artefacts.  Further details of this technique are provided in Section 4.1 of \citet{fletcher_impact_2011}. We likewise adopt and re-analyze this corrected version of the T-ReCS spectra in this work.

In inverting the T-ReCS spectra, \citet{fletcher_impact_2011} used the multi emission angle observations of the same location to improve the vertical sensitivity of retrieved atmospheric parameters.  In this work, we instead adopt a single, average spectrum such that the treatment of spectra for the SL9 and Wesley impacts are similar as possible. Nominally, spectra of the impact and background atmosphere were computed by coadding individual spectra recorded at respective longitude ranges of 299 -307$^\circ$W and 316-325$^\circ$W, as shown in Figure \ref{fig:trecs_spx}a-b.  While using all available spectra within these longitude ranges results in a higher signal-to-noise ratio, they capture the impact site and background atmosphere at very different mean emission angles.  For a direct comparison of the spectra between impact site and background atmosphere, we repeated the coaddition of spectra over the same longitude ranges but omitted a subset of individual spectra such that the impact and background spectra were similar in mean emission angle.  These are shown in Figure \ref{fig:trecs_spx}c-d and the fractional residual between them are shown in Figure \ref{fig:trecs_spx}e-f. 

The atmosphere at the site of the Wesley impact is enhanced in emission from 8.8 to 12 $\upmu$m and 17 – 19 $\upmu$m in comparison to the unperturbed atmosphere.  Again, the 8.8-to-12 $\upmu$m feature was nominally attributed to some mixture of stratospheric NH$_3$ gas (in emission) and non-gaseous material.  The 17 – 19 $\upmu$m feature was also attributed to non-gaseous material.  In Section \ref{sec:mineral_modelling}, we adopt the fractional residual computed from spectra similar in emission angle (Figure \ref{fig:trecs_spx}c-f) as an emissivity spectrum of the impact region. 

For the purpose of comparing relative wavelength dependence between the impacts, the fractional residuals for each dataset were normalized, as shown in Figure \ref{fig:compare_all}.  This higlights two main contrasts between the SL9 impacts and the Wesley impact.  First, the SL9 impact sites exhibited enhanced 7 – 8 $\upmu$m stratospheric emissions whereas the Wesley impact site was not enhanced in CH$_4$ emissions outside of uncertainty, which is in agreement with previous work (e.g. \citealt{orton_2011},\citealt{fletcher_impact_2011}).  Second, the broad N-band feature extends from 8.8 to 13 $\upmu$m whereas for SL9, the broad feature extends to shorter wavelengths of $\sim$8.5 $\upmu$m.  As we demonstrate in Section \ref{sec:mineral_modelling}, the N-band feature extends to shorter wavelengths in the SL9 impact due to the presence of obsidian (glassy silica) and its absence from the Wesley impact spectrum.

\section{Analysis \& Results}\label{sec:analysis}

We performed two separate analyses to determine the species responsible for the observed emission features in the SL9 and Wesley impact sites.  Each analysis has advantages and disadvantages over the other. First, the fractional residual between the spectra of the impact region and a region away from the impact was calculated and then normalized.  The normalized fractional residual spectrum was then modelled by performing a least-squares search over a grid of candidate mineral species with varying abundances.  In this part of the analysis, gaseous NH$_3$ in Jupiter's stratosphere was also treated as a ``mineral" and included in the model grid search.  The advantage of this method is that gaseous NH$_3$ and non-gaseous mineral species are constrained simultaneously.  The disadvantage is that the calculation of the fractional residual assumes the temperature of the impact region and off-impact region are similar and were measured at a similar emission angle.  In the Wesley impact, negligible stratospheric heating was observed over the impact region (Figure \ref{fig:trecs_spx}, \citealt{fletcher_impact_2011}) and both regions were sampled at a range of emission angles, which allowed spectra similar in emission angle to be calculated.  However, for SL9, strong stratospheric heating was observed over the impact regions and it was impossible to sample the impact and off-impact region at similar emission angles.  Thus, the assumption of similar temperature and emission angle is valid for the Wesley impact but is a poor assumption for the SL9 impact.   This analysis is detailed further in Section \ref{sec:mineral_modelling}. 

In the second approach, we inverted the spectra using the NEMESIS radiative transfer code \citep{irwin_2008}.  The vertical profiles of temperature and stratospheric hydrocarbons were allowed to vary and retrievals were performed over a grid of NH$_3$ vertical profiles.  The combination of parameters producing a synthetic spectrum that minimized the goodness-of-fit was interpreted to represent the gaseous component of the spectrum.   Wavelengths where the synthetic spectra did not adequately fit the observed spectra were interpreted to result from non-gaseous emission from impactor material and the wavelength dependence of the non-gaseous emission was compared with the absorptance spectra of several candidate mineral species.  Unlike \citet{fletcher_impact_2011}, we did not attempt to include mineral species in the radiative transfer inversion to avoid introducing degenerate parameters such as aerosol size distribution and vertical profiles to an already large parameter space.  The advantage of this method is that the effects of temperature and emission angle on the planet are fully characterized, a range of vertical profiles of NH$_3$ can be tested, and absolute temperatures and NH$_3$ concentrations are constrained.  The disadvantage is that the inversion could misinterpret non-gaseous emission as gaseous emission.  This analysis is detailed further in Section \ref{sec:inversions}.  

\subsection{Mineralogical dust modelling}\label{sec:mineral_modelling}

Mineralogical spectral modeling was performed by adopting a collection of candidate mineralogical species, performing a grid search to find a combination of species that produced a model emissivity spectrum that minimized the goodness-of-fit ($\chi^2/n$, where $n$ is the number of degrees of freedom) to the emergent emissivity spectra of both impacts (Figure \ref{fig:palomar_spx} and \ref{fig:trecs_spx}).  For an optically thin refractory residue, a composite spectrum can be written as the combination of the contributions from individual species \citep{lisse_2009}, as in Equation \ref{eq:Fv}

%Given SL9 was a known comet and the suggestion in previous studies that the 2009 Wesley impactor was an asteroid, the expected differences in mineralogy include a preponderance of amorphous olivine and pyroxene, plus water ice and amorphous carbon, in the D/SL9 cometary impact, versus a preponderance of crystalline olivine and phyllosilicates plus silicas for the Wesley 2009 putatively “asteroidal” impact case. 

\begin{equation}\label{eq:Fv}
F_v = \frac{B_\nu [T_{atm}]}{\Delta^2} \sum_{i}  \int_{a_{min}}^{a_{max}} Q_{\mbox{abs},i} (a,\lambda) \pi a^2 \frac{dn_i (r)}{da} da
\end{equation}

where $\Delta$ is the distance between the observer and the dust, $B_\nu$ is the Planck function at wavelength $\lambda$ and atmospheric temperature $T_{atm}$.  $a$ is the particle radius, $r$ is the distance from the Sun, $Q_{abs,i}$ is the emission efficiency of the i$^{th}$ species and $dn/da$ is the differential particle size distribution as a function of radius, $a$. The emitted flux depends on the composition (location of spectral features) and the particle size (feature to continuum contrast). The particle size distribution was assumed to be a power law or power law reduced at small particle sizes to model radiation-pressure blowout effects.  For the candidate mineralogical species, we nominally considered $\sim$100 species that have previously been used to fit the mid-infrared spectra of dust produced by comets, asteroids and jovian trojans in our solar system and by exocomets, exoasteroids, and aggregating planetesimals in other nearby star systems \citep{lisse_2006,lisse_2007a,lisse_2007b,lisse_2008,lisse_2009,lisse_2012,lisse_2017}, including silica-rich hyper velocity impact systems like HD172555 \citep{lisse_2009,lisse_2020}.  These species include multiple amorphous silicates with olivine-like and pyroxene-like composition; multiple ferromagnesian silicates (forsterite, fayalite, clino- and ortho-pyroxene, augite, anorthite, bronzite, diopside, and ferrosilite); silicas, both amorphous and crystalline (obsidian, tektites, quartz, cristobalite, tridymite); phyllosilicates (such as saponite, serpentine, smectite, montmorillonite, and chlorite); sulfates (such as gypsum, ferrosulfate, and magnesium sulfate); oxides (including various aluminas, spinels, hibonite, magnetite, and hematite); Mg/ Fe sulfides  (including pyrrohtite, troilite, pyrite, and ningerite); carbonate minerals (including calcite, aragonite, dolomite, magnesite, and siderite); water-ice, clean and with carbon dioxide, carbon monoxide, methane, and ammonia clathrates; carbon dioxide ice; graphitic and amorphous carbon; and the neutral and ionized polycyclic aromatic hydrocarbon (PAH) emission models of \citet{draine_2007}.   The Supplementary material of \citet{lisse_2006} details the sources of absorption spectra of the aforementioned species.  However, for both impacts in this study, the wavelengths at which non-gaseous spectral features were and were not observed (Figure \ref{fig:compare_all}) ruled out the large majority of these nominal species.  Ultimately, we only considered the mineralogical species listed in Table \ref{tab:mineral}.  
\begin{figure}[ht!]
\begin{center}
\includegraphics[width=0.45\textwidth]{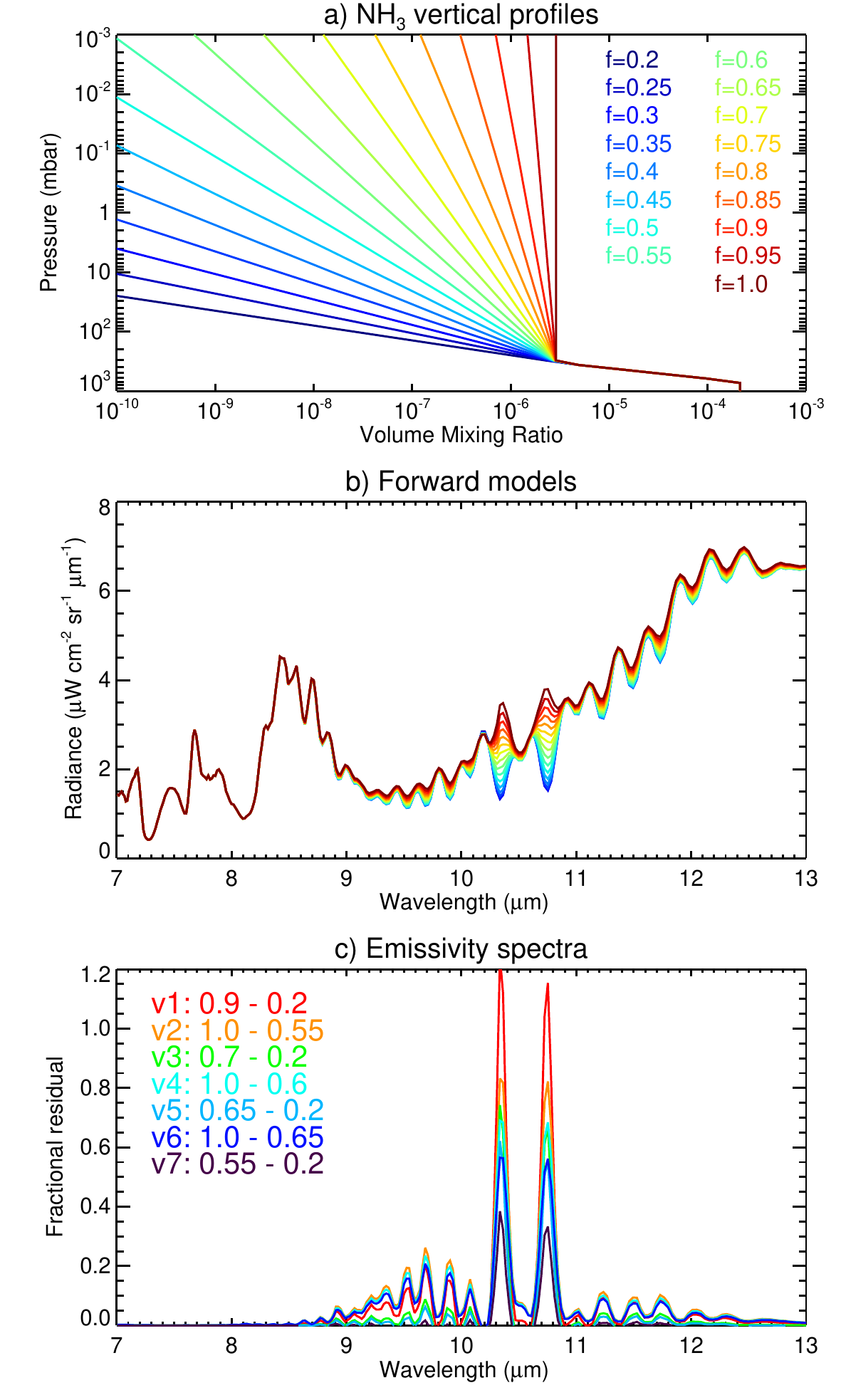}
\caption{(a) A family of NH$_3$ vertical profiles with varying fractional scale heights, as indicated in the legend, to parameterize stratospheric NH$_3$.  Panel (b) shows the corresponding forward model spectra at T-ReCS/SC-10 spectral resolution using the same color conversion.  Panel (c) shows seven NH$_3$ emissivity spectra computed by finding the fractional residual between two forward model spectra, as detailed by the legend.  For example, spectrum `v1' was computed by finding the residual between the forward model spectra using fractional scale heights of 0.9 and 0.2. }
\label{fig:compare_nh3}
\end{center}
\end{figure}

Since gaseous NH$_3$ lofted into the stratosphere has a significant contribution to the observed spectral features, we also included and treated gaseous NH$_3$ as a ``mineral'' in our spectral fitting. In order to compute a emissivity spectra of NH$_3$, we forward modelled spectra of Jupiter with a range of vertical NH$_3$ profiles and computed their fractional difference.  Section \ref{sec:inversions} provides further details of the radiative transfer model and atmospheric model of Jupiter used to compute forward model spectra.  The nominal, vertical NH$_3$ profile was retrieved from Cassini-CIRS spectra at 45$^\circ$S on Jupiter and assumes NH$_3$ is well mixed with a volume mixing ratio of 360 ppmv at pressures higher than 890 mbar, and then decreases in abundance at lower pressures according to a fractional scale height of 0.21 (see \ref{sec:cirs_retrievals} for further details).  In order to parameterize NH$_3$ gas lofted into the stratosphere, we adopted the approach presented by \citet{fletcher_impact_2011}, where the NH$_3$ profile, $q$, is set to the CIRS-derived profile, $q_{CIRS} (p)$ at pressures higher than a cutoff pressure, $p_0$, and then a decrease in abundance with altitude quantified by a fractional scale height, $f$.  This is quantified in Equation \ref{eq:nh3} below.

\begin{equation} \label{eq:nh3}
\begin{split}
q (p \ge p_0) & = q_{CIRS} (p) \\ 
q (p < p_0) & = q_0 \left(\frac{p}{p_0} \right) ^{\frac{1-f}{f}}
\end{split}
\end{equation}

\begin{table*}[!t]
\centering  % used for centering table
\begin{tabular}{>{\centering\arraybackslash} m{4.0cm} |>{\centering\arraybackslash} m{1.5cm}  >{\centering\arraybackslash} m{1.5cm} |>{\centering\arraybackslash} m{1.5cm}>{\centering\arraybackslash} m{1.5cm} } 
Species & \multicolumn{2}{c}{SL9 G (1994)} & \multicolumn{2}{c}{Wesley (2009)} \\
	      & \% & $\Delta \chi^2$ & \% & $\Delta \chi^2$ \\
\firsthline 
Amorphous Olivine & 0.3 & 16.6 & 0.65 & 27.3 \\
Amorphous Pyroxene & 0 &0 & 0 & 0\\
Obsidian & 0.32 & 1.3 & 0 & 0\\
Fosterite & 0.03 & 0.04 & 0.01& 0\\
Fayalite & 0 & 0 &0  & 0\\
Diopside & 0.08 & 0.1 & 0 & 0\\
Ferrosite &0 &0 & 0.07& 0.06 \\
Ortho Enstatite & 0 & 0& 0.02& 0.01 \\
Australite Tektite & 0 & 0& 0& 0 \\
Bediasite Tektite& 0.03 & 0.2& 0& 0\\
MgFeS &0 & 0&0 & 0\\
PAH & 0 & 0& 0& 0\\
NH$_3$ (g) & 0.2 & 1.0& 0.24 & 0.7\\
Amorphous Carbon & 0 & 0& 0 & 0\\
H$_2$O (s) & 0.02 & 0.1& 0 & 0\\
\hline
$\chi^2/n$ & \multicolumn{2}{c}{1.1} &\multicolumn{2}{c}{0.91} \\
\lasthline
\end{tabular}
\caption{The list of species included in the spectral fitting, the derived proportion of each mineral (\%) and the change in the reduced $\chi^2$ ($\Delta \chi^2$) when that species is omitted from the spectral fitting.   Results for the SL9 G impact site are shown in the middle column and for the Wesley impact in the right-hand column.  }
\label{tab:mineral}
\end{table*}
\begin{figure*}[ht!]
\begin{center}
\includegraphics[width=0.91\textwidth]{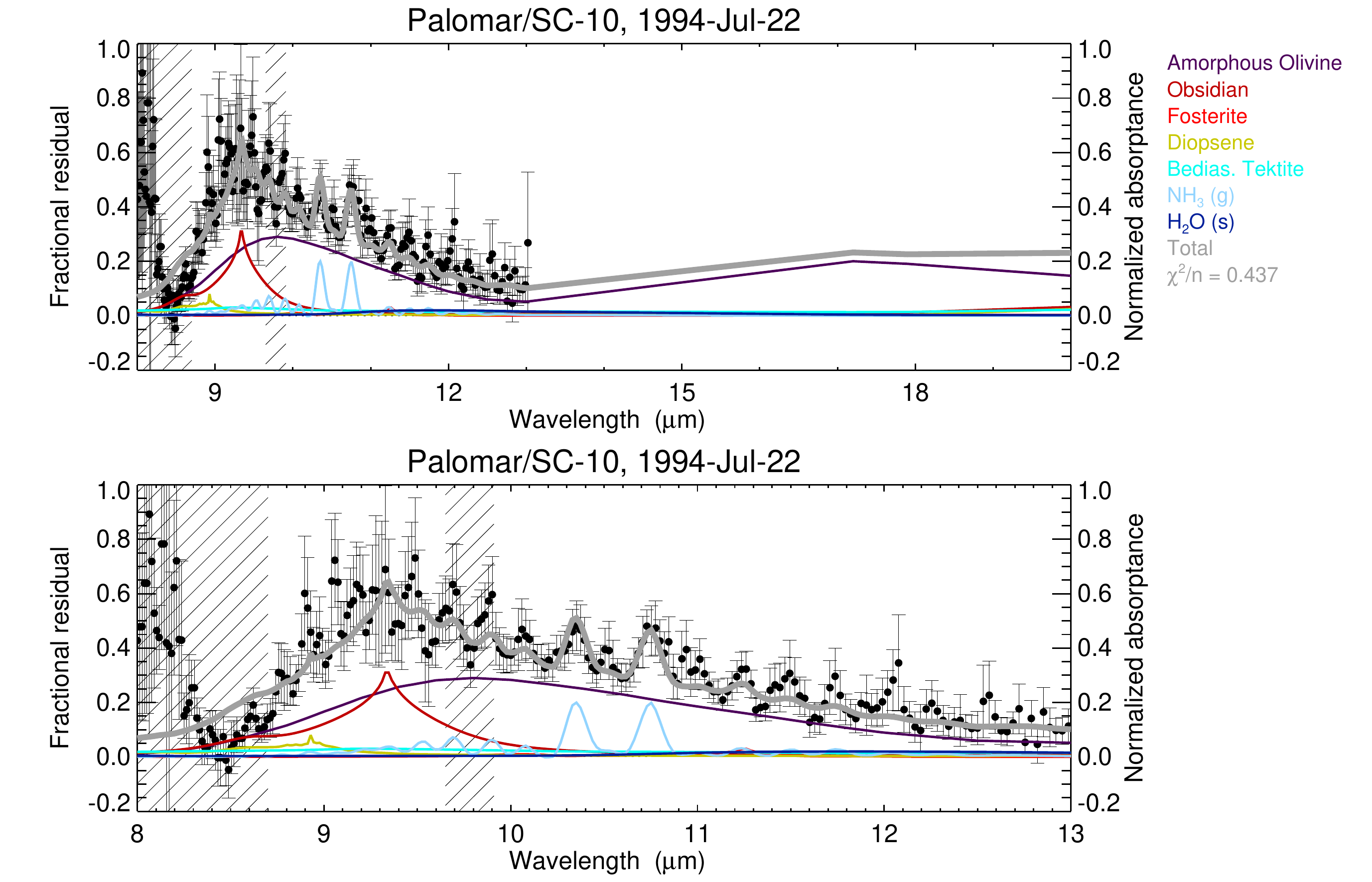}
\caption{The observed fractional residual of SL9 Impact G using Palomar/SC-10 spectroscopy and the best-fitting total mineralogical spectrum (solid grey).  The top panel shows the full wavelength range, the lower panel focuses on the 8 - 13 $\upmu$m range. The individual species that contribute to the total mineralogical curve (see Table \ref{tab:mineral}) are shown and colored according to the legend.   Areas covered by 45$^\circ$-lines denote the wavelengths omitted in performing the mineralogical fitting due to contamination by telluric O$_3$ or jovian, stratospheric (gaseous) CH$_4$ emission). }
\label{fig:compare_palomar_mineral}
\end{center}
\end{figure*}

\begin{figure*}[ht!]
\begin{center}
\includegraphics[width=0.91\textwidth]{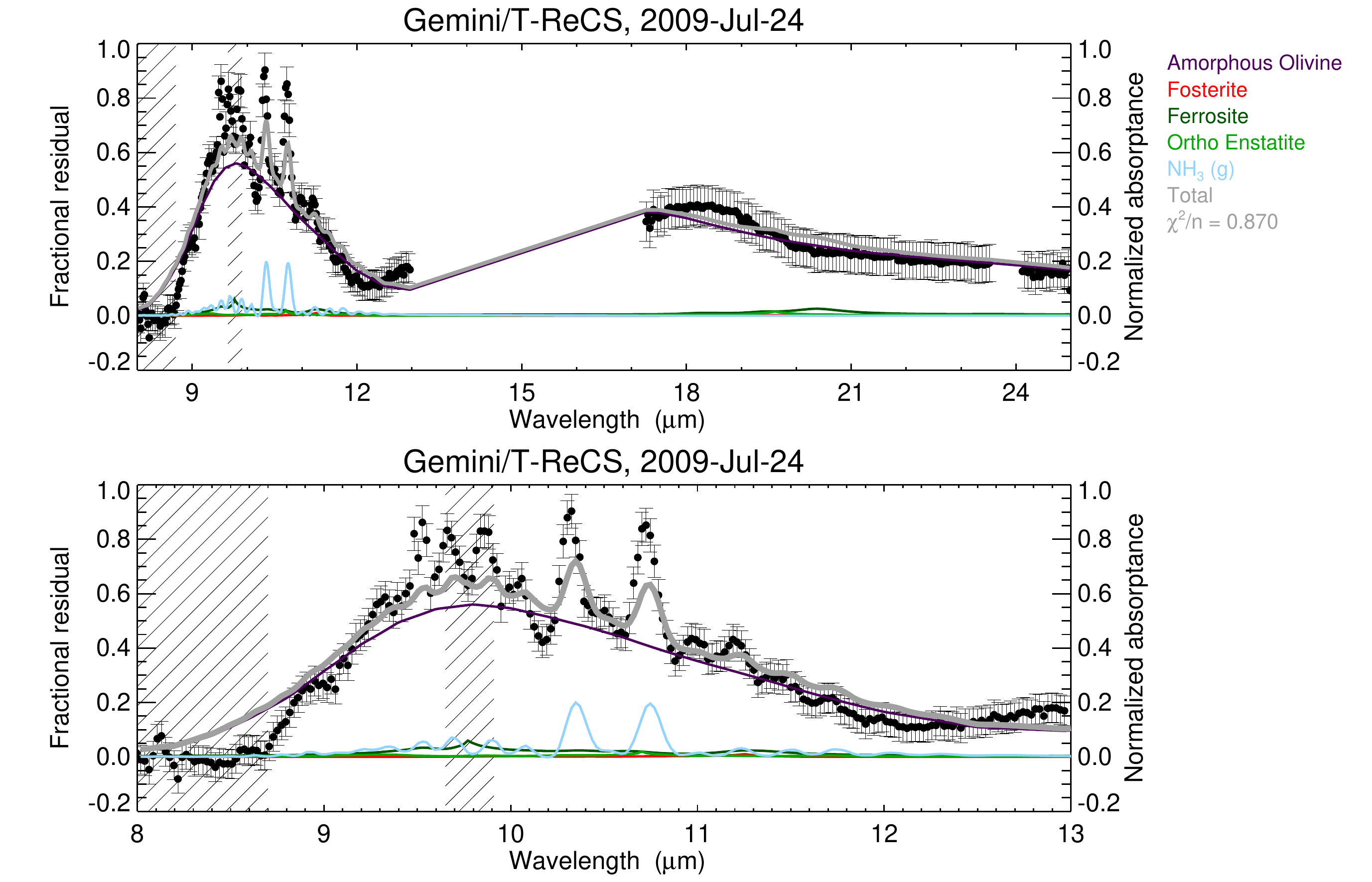}
\caption{The observed fractional residual of the Wesley impact using Gemini-S/T-ReCS spectroscopy and the best-fitting total mineralogical spectrum (solid grey).  The top panel shows the full wavelength range, the lower panel focuses on the 8 - 13 $\upmu$m range. The individual species that contribute to the total mineralogical curve (see Table \ref{tab:mineral}) are shown and colored according to the legend.   Areas covered by 45$^\circ$-lines denote the wavelengths omitted in performing the mineralogical fitting due to contamination by telluric O$_3$ or jovian, stratospheric (gaseous) CH$_4$ emission). }
\label{fig:compare_trecs_mineral}
\end{center}
\end{figure*}

 $q_0$ is the abundance of NH$_3$ at pressure, $p_0$, $f$ is a fractional scale height (between 0 and 1).  For the purposes of computing emissivity spectra of NH$_3$, we assumed $p_0$ = 330 mbar and varied $f$ between 0.2 and 1.0 in increments of 0.05.  However, we note to readers that the vertical profile of NH$_3$ in the SL9 impact regions may be more complex than the power law parameterization we adopted here.

 Figure \ref{fig:compare_nh3}a, b show the resulting vertical profiles of NH$_3$ and corresponding forward model spectra at T-ReCS/SC-10 spectral resolution.  We computed a range of NH$_3$ emissivity spectra by calculating the fractional residual between different combinations of forward model spectra, as shown in Figure \ref{fig:compare_nh3}c.  The goal was to produce a range of spectra with varying strengths of the stronger, 10.3- and 10.8-$\upmu$m lines with respect to the weaker lines between 9 - 10 $\upmu$m and 11 - 12 $\upmu$m.  In modeling the emission spectra of the SL9 and Wesley impacts, each emissivity spectrum was adopted in turn, and scaled by a factor at all wavelengths.  We found that emissivity spectrum v2 (the fractional residual between the forward models using fractional scale heights of 1.0 and 0.55) yielded the best fits to the Palomar/SC-10 spectra of the SL9 G impact, whereas v6 yielded the best fits to the Gemini/T-ReCS spectra of the Wesley impact.

After testing the spectral mineralogical fitting with several different size distributions, we found that a single power law of $dn/da$ $\sim$ $a^{-4.20}$ optimized the fit to the Palomar/SC-10 spectra of the SL9 G impact whereas a power law index of -4.30 best reproduced the Gemini-S/T-ReCS spectra of the Wesley impact.   Both are extremely steep, as expected for impact-produced dust populations dominated by small, sub-micron- to micron-sized particles \citep{lisse_2009,takasawa_2011,johnson_2012}. While T-ReCS spectra of the Wesley impact recorded both the N and Q band, Palomar spectra of the SL9 impact only recorded the N band.  In order for the analysis to be as consistent as possible for both impacts, we included the IRTF-MIRAC broadband 17.2-, 17.9 and 20.8-$\upmu$m measurements of the SL9 G impact (Figures \ref{fig:fr_mirac}) in the fitting of the Palomar spectra.  The MIRAC measurements were scaled by a factor of 4 such that the difference in fractional residual between the Palomar N-band and MIRAC Q-band measurements was similar as the difference in fractional residual between the MIRAC N- and Q-band measurements.   

Finally, for both impacts, we omitted wavelengths 9.7 - 9.9 $\upmu$m from the spectral fitting since these wavelengths are obscured by telluric ozone.  For Palomar/SC-10 spectra of SL9, wavelengths shorter than $\sim$8.5 $\upmu$m contain strong stratospheric CH$_4$ emission, which we did not want to be interpreted as non-gaseous emission by the spectral fitting.  For the Gemini/T-ReCS spectrum of the Wesley impact, we found that including the 8.0 -  8.7 $\upmu$m drove unphysical solutions for mineralogical compositions.   In order to analyze both impacts/observations as consistently as possible, we chose to omit wavelengths shorter than 8.7 $\upmu$m in the fitting of both observations.

Table \ref{tab:mineral} details the best-fitting compositions of minerals for each impact and Figures \ref{fig:compare_palomar_mineral} and \ref{fig:compare_trecs_mineral} compare the fractional residuals of both impacts with the best-fitting mineralogical composition.  For the SL9 G impact, we found that the main components producing significant emissivities were amorphous olivine, silica (in the form of obsidian), and gaseous NH$_3$ in the stratosphere.  A small amount of the water ice, diopside, and forsterite typically found in comets was present in the D/SL9 case, but their absence from the synthetic spectrum had only a small, statistically-insignificant effect on the reduced $\chi^2$.  We found no evidence of amorphous pyroxene in the D/SL9 case, which was initially unexpected since comets typically have a $\sim$1:1 ratio of pyroxenaceous to olivinaceous species (e.g. \citealt{lisse_2006,lisse_2007b}).  As we discuss in Section \ref{sec:discuss}, we suggest that all the cometary pyroxene was converted into silicas by the strong heating ($>$ 1000 K) associated with the impact.  A similar finding was seen in the HD172555 exodisk system by \citet{lisse_2009}.

By contrast, the Wesley impact emission appears composed of almost pure amorphous olivine, with some ferromagnesian pyroxenes mixed in at an $\sim$8:1 ratio, together with gaseous NH$_3$. No emission due to silica is apparent in our new reduction and spectral analysis, a marked difference from the findings of \citet{fletcher_impact_2011}. Without any evidence of silica, it is impossible to understand this mixture of materials being formed at high temperatures.  In order to explain the conversion of the usually dominant crystalline olivine into amorphous olivine, we suggest that the Wesley residue was formed at low temperatures and high pressures, as discussed further in Section \ref{sec:discuss}.  

Figures \ref{fig:compare_palomar_mineral} and \ref{fig:compare_trecs_mineral} also demonstrate the \textit{a posteriori} fit of the best-fitting mineralogical composition to the 8.0 - 8.7 $\upmu$m spectral region that was omitted from the least-squares search.  For SL9, the fit to the Palomar/SC-10 spectrum between 8.3 - 8.7 $\upmu$m is adequate (within uncertainty).  The fit is expectedly poorer from 8 - 8.3 $\upmu$m, which captures stratospheric CH$_4$ emission that was not included in the grid search.  For the Wesley impact, the fit of the best-fitting mineralogical composition does not adequately fit the 8.3 - 8.7 $\upmu$m region.  This suggests a missing (and currently unknown) source of opacity from 8.3 - 8.7 $\upmu$m in the impact region, which is in absorption rather than in emission.  

%Whether these pressures were achieved due to the shock of impact, or due to the pressures achieved at depth inside a large differentiated body, is not clear. We currently favor the impact-induced shock amorphization explanation, as all reports have the Wesley impactor as being relatively small in size (< 10 km radius), too small to thoroughly pressure-alter its olivine throughout; also, such an explanation would require removal of the body’s low pressure, near-surface crystalline olivine without heating the high pressure phase enough to produce silica. This still belabors the question of how a body impinging on Jupiter with ~65 km/sec (Jupiter’s escape speed) relative velocity and a huge specific energy per kg of material can manage to NOT melt and transform its matter, just the way D/SL9 was seen to create large amounts of glassy silica (a material not found in comets). A remote possibility is that the Wesley 2009 impactor “came in” at an extremely shallow angle, i.e. almost parallel to the local atmospheric surface, allowing material to spall off the impactor at relatively low temperatures. If we take the velocity of impact required to convert silicates into silicas as ~10 km/sec (\citealt{lisse_2009} and references therein), then this would require impact angles of less than tan$^-1(10/65)$ $\sim$ 9$^\circ$ with respect to the local horizontal in order to impart the impactor kinetic energy to the body slowly enough that it could have a chance to retain all its silicates.

\subsection{Radiative transfer inversions}\label{sec:inversions}

We adopt the NEMESIS (Non-linear optimal Estimator for MultivariatE spectral analysis, \citet{irwin_2008}) radiative transfer code to perform inversions of the data.  This code has been used extensively in previous investigations \citep{orton_2011,fletcher_2009,fletcher_impact_2010,fletcher_impact_2011}.  Upper-tropospheric temperatures were sensed using H$_2$-related collision-induced absorption (CIA) continuum that dominates wavelengths in the 17-25 $\upmu$m range, with coefficients derived by \textit{ab initio} models for H$_2$-H$_2$ \citep{fletcher_dimers_2018}, H$_2$-He \citep{birnbaum_1996}, and H$_2$-CH$_4$  \citep{borysow_1986}.   The spectroscopic line information for CH$_4$ and its isotopologues, C$_2$H$_2$, C$_2$H$_4$, C$_2$H$_6$, NH$_3$ and PH$_3$, which are the relevant and dominant gaseous species at the wavelengths of this study, were adopted from Supplementary Table 1 of \citet{fletcher_poles_2018}.  

For computational efficiency, we chose to perform forward models and inversions using the correlated-k method, where absorption coefficients of all relevant gases within a wavenumber are sorted in order of line strength and the distributions of line strengths, or k distribution, is computed \citep{correlated_k,rodgers_2000}.  In order to model both Gemini/ T-ReCS and Palomar/SC-10 spectra, the aforementioned spectroscopic line information was convolved with a triangular line function with a FWHM of $\Delta \nu $= 10 cm$^{-1}$ and $\Delta \nu $ = 5.55 cm$^{-1}$ to model the N-band (7 – 13 $\upmu$m) and Q-band (17 – 25 $\upmu$m), respectively.  K-distributions in each band were calculated and concatenated such that the N- and Q-band spectra could be modelled simultaneously.   

\subsubsection{Modeling Palomar/SC-10 spectra}\label{sec:nemesis_sc10}

\begin{figure*}[t!]
\begin{center}
\includegraphics[width=0.85\textwidth]{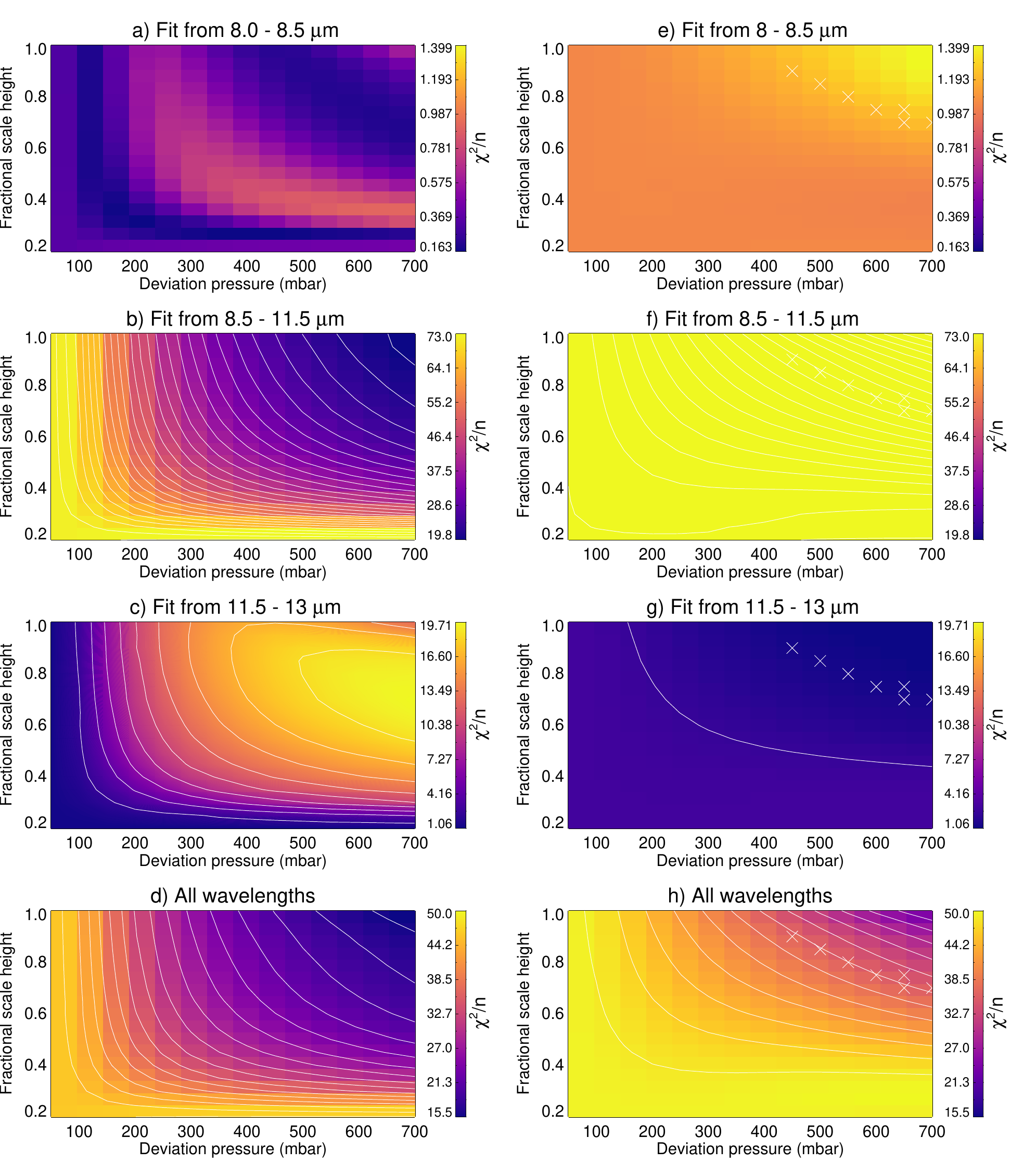}
\caption{Reduced $\chi^2$ values in fitting the Palomar/SC-10 spectra as a function of deviation pressure, $p_0$ and fractional scale height of the NH$_3$ (gas) profile.  The left column shows results derived from inverting the full 8 – 13 $\upmu$m spectral range, the right column shows results derived from inverting the 8 – 8.5 and 11.5 – 13 $\upmu$m spectral range (and radiances in the 8.5 – 11.5 $\upmu$m range were forward modelled).  are shown in the left column, Results are shown  at wavelengths of 8 – 8.5 $\upmu$m (R branch of CH$_4$ emission, continuum), 8.5 – 11.5 $\upmu$m (NH$_3$ and non-gaseous emission), 11.5 – 13 $\upmu$m (C$_2$H$_6$ emission) and all wavelengths.  White lines denote $\chi^2/n$  increments of 2.   White crosses denote model results that fit the spectra within absolute $\chi_{min}^2$ + 2.3 (the 1-$\sigma$ confidence level) and reproduce the observed strength of NH$_3$ emission at 10.36 and 10.75 $\upmu$m within uncertainty.  }
\label{fig:palomar_grid_results}
\end{center}
\end{figure*}

Adopting a initial guess or \textit{a priori} atmosphere derived from Cassini-CIRS (Composite Infrared Spectrometer) measurements of 40$^\circ$S in 2001 (see \ref{sec:cirs_retrievals}), we fit the observed Palomar/SC-10 8–13 $\upmu$m spectra of the G impact site (Figure \ref{fig:palomar_spx}) by allowing the vertical profiles of temperature, NH$_3$, the abundance of tropospheric aerosol and C$_2$H$_4$ to vary. The tropospheric aerosol and stratospheric C$_2$H$_4$ abundances were parameterized by retrieving a scale factor applied to all altitudes of their respective CIRS-derived profiles.  Temperature was allowed to vary continuously at all altitudes with wavelengths 8.2 – 8.5 $\upmu$m providing sensitivity to the upper troposphere (1– 5 mbar) and CH$_4$ (8 - 8.1 $\upmu$m) and C$_2$H$_6$ emission (11.3 – 13.0 $\upmu$m) providing sensitivity to the lower stratosphere (50 mbar $<$ p $<$ 0.1 mbar).  Outside of the range of sensitivity, retrieved temperatures tend back to \textit{a priori} (CIRS-derived) values.  As in Section \ref{sec:mineral_modelling}, we parameterized the vertical profile of NH$_3$ using the approach presented in \citet{fletcher_impact_2011}.   A 2-dimensional model grid was computed with 13 $p_0$ values ranging from 50 mbar to 7 mbar in increments of 50 mbar and fractional scale heights from 0.2 to 1 in steps of 0.05.   The vertical profiles of NH$_3$ of the model grid were adopted in turn, fixed and the SC-10 spectrum of the SL9 G impact was inverted.  In previous work analyzing observations of the SL9 impacts, a range of parameterizations of the stratospheric NH$_3$ vertical profile were adopted, including a step profile (e.g. \citealt{fast_2002}) and a double-peaked profile (e.g. \citealt{griffith_1997}).  However, the observations analyzed in those studies were of high spectral resolving powers ($\sim$10$^4$ to 10$^7$) and therefore better resolve the vertical structure in the impact regions compared to the $\sim$10$^2$ resolving powers analyzed in this work. We therefore chose to parameterize the vertical profile of NH$_3$ using a simple power law index (two parameters) to avoid a highly degenerate parameter space and to be consistent with our parameterization of the NH$_3$ vertical profile in analyzing T-ReCS spectra of the Wesley impact.

At mid-infrared wavelengths, the magnitude of stratospheric hydrocarbon emissions is modulated both by the vertical temperature profile and the abundance of the emitting species.   In previous work (e.g. \citealt{nixon_meridional_2007,nixon_2011,fletcher_2016,sinclair_2020b}), stratospheric temperatures are normally derived from fitting CH$_4$ emission at 7.2 – 8.5 $\upmu$m since the vertical profile of CH$_4$ is generally assumed to be horizontally homogeneous in the lower stratosphere.  However, in this work, there are only a few spectral points capturing CH$_4$ emission in this wavelength range and with a poorer signal-to-noise ratio (SNR) due to telluric obscuration.  As a result, the vertical temperature profile is largely constrained from fitting C$_2$H$_6$ emission, since it is sampled by a greater number of spectral points and at a higher SNR.  Our analysis assumes the CIRS-derived C$_2$H$_6$ profile at 45$^\circ$S derived from Cassini-CIRS measurements recorded during the 2001 Jupiter flyby, however, stratospheric abundances of C$_2$H$_6$ do vary temporally due radiative forcing and dynamics (e.g. \citealt{melin_2018}, \citealt{hue_2018}).  Thus,  absolute stratospheric temperatures and abundances should be interpreted with caution. 

\begin{figure}[t!]
\begin{center}
\includegraphics[width=0.48\textwidth]{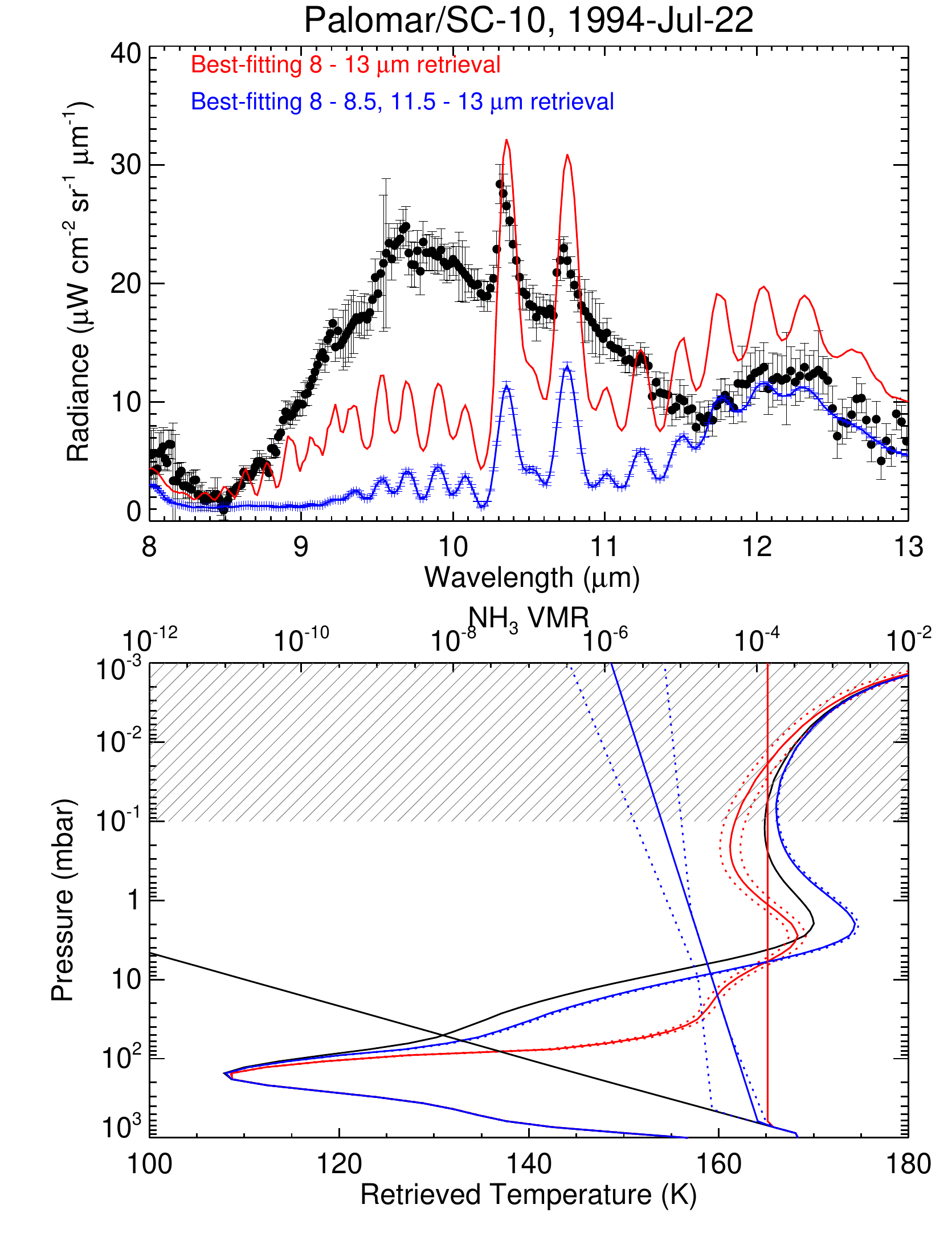}
\caption{The top panel shows the observed Palomar/SC-10 spectrum (black points with error bars) and model spectra are shown as solid, colored lines.  The red spectrum corresponds to the “best-fitting” result when inverting the full 8 – 13 $\upmu$m range, which produces too much NH$_3$ emission at wavelengths longer than 11 $\upmu$m.   The blue spectrum is the best-fitting result when inverting only the 8 – 8.5 and 11.5 – 13 $\upmu$m spectral range and that produces NH$_3$ emission features of a similar strength as observed within uncertainty - see the text for further details.   The blue synthetic spectrum is considered to represent the component of the observed spectrum formed by gaseous emission, as detailed further in the text.  The bottom panel shows the corresponding temperature and NH$_3$ profiles (solid lines) associated with the synthetic spectra.  Dotted lines represent 1-$\sigma$ uncertainty. The black, solid lines denote the \textit{a priori} (CIRS-derived) temperature and NH$_3$ profiles adopted in inverting the spectra.  Hashed regions at pressures lower than 0.1 mbar indicate where the observations have little/no sensitivity.}
\label{fig:palomar_example_fits}
\end{center}
\end{figure}

Figure \ref{fig:palomar_grid_results}a-d shows the reduced $\chi^2$ fits to the Palomar spectrum at 8.0 – 8.5 $\upmu$m (capturing continuum and stratospheric CH$_4$ emission), 8.5 – 11.5 $\upmu$m (predominantly NH$_3$ gaseous emission and non-gaseous impact residue ), 11.5 – 13.0 $\upmu$m (continuum and stratospheric C$_2$H$_6$ emission) and over all wavelengths.   We find there is no model atmosphere within the described model grid that fits all wavelengths adequately, i.e. with $\chi^2/n$  $\sim$ 1.  While the 8.5 – 11.5 $\upmu$m spectral range is relatively better fit ($\chi^2/n$   $\sim$ 20) with models assuming higher values of $p_0$ and $f$ (i.e. higher NH$_3$ abundances over a larger vertical range of atmosphere), the same models produce too much NH$_3$ emission and therefore a poorer fit in the 11.5 – 13.0 $\upmu$m spectral range. This is also demonstrated in Figure \ref{fig:palomar_example_fits}.  We attribute the inability to fit all wavelengths with the same temperature, aerosol and ammonia profile due to the presence of significant non-gaseous emission predominantly in the 8.5 – 11.5 $\upmu$m spectral range.

Given the inability to fit all wavelengths of the spectrum due to non-gaseous emission, we instead modelled the spectrum in the following way.  The retrievals of temperature and aerosol abundance were repeated across the same model grid of NH$_3$ profiles (see above) but using only the 8 – 8.5 and 11.5 – 13 $\upmu$m spectral ranges.  These ranges were chosen because they capture CH$_4$ and C$_2$H$_6$ emission features, and also correspond to wavelength ranges where the candidate mineralogical species (see Section \ref{sec:mineral_modelling}) have significantly weaker absorption features.  The retrieved atmosphere was then forward-modelled across the entire 8 – 13 $\upmu$m spectral range and the reduced $\chi^2/n$  values were calculated.  The goodness-of-fit values for these results are shown in \ref{fig:palomar_grid_results}e-h. In omitting the 8.5 – 11.5 $\upmu$m spectral range, the retrieval is able to adequately fit ($\chi^2/n \sim 1$) the 8 – 8.5 and 11.5 – 13 $\upmu$m spectral range with the same atmosphere. 

In order to calculate the component of the spectrum produced by gaseous emission, we searched the model grid results for results that satisfied the following criteria.  First, the synthetic spectrum must fit the observed spectrum in the 8.0 – 8.5 and 11.5 – 13 $\upmu$m ranges with an absolute $\chi^2$ value of less than $\chi_{min}^2$ + 2.3, which denotes the 1-$\sigma$ confidence level \citep{press_1992} when varying two parameters.  Second, the synthetic spectrum must reproduce the line-to-continuum radiance difference of the strong NH$_3$ emission features at 10.36 and 10.75 $\upmu$m within uncertainty on the radiance. The models that satisfy these criteria are shown as white crosses in Figure \ref{fig:palomar_grid_results}e-h.  A model atmosphere with the NH$_3$ profile deviating from the CIRS-derived profile at 650 mbar and with a fractional scale height of 0.75 yields the best-fitting spectrum that satisfy these criteria. The spectrum associated with this model was adopted as the gaseous emission spectrum.  The uncertainty of the gaseous component of the spectrum was determined by calculating the  standard deviation of the synthetic spectra that satisfied the aforementioned criteria.  Similarly, at each atmospheric level, the standard deviation in NH$_3$ values from the range of the model atmosphere that satisfied the aforementioned criteria were calculated and adopted as their respective uncertainties.  We performed a similar calculation to determine the uncertainty on temperature but found that the standard deviation was smaller than the uncertainty on temperature from the best-fitting retrieval and so the latter was adopted as the uncertainty on temperature.  The gaseous emission spectrum and the associated vertical profiles of temperature and NH$_3$ and their uncertainties are shown in Figure \ref{fig:palomar_example_fits}. 
\begin{figure*}[t!]
\begin{center}
\includegraphics[width=0.9\textwidth]{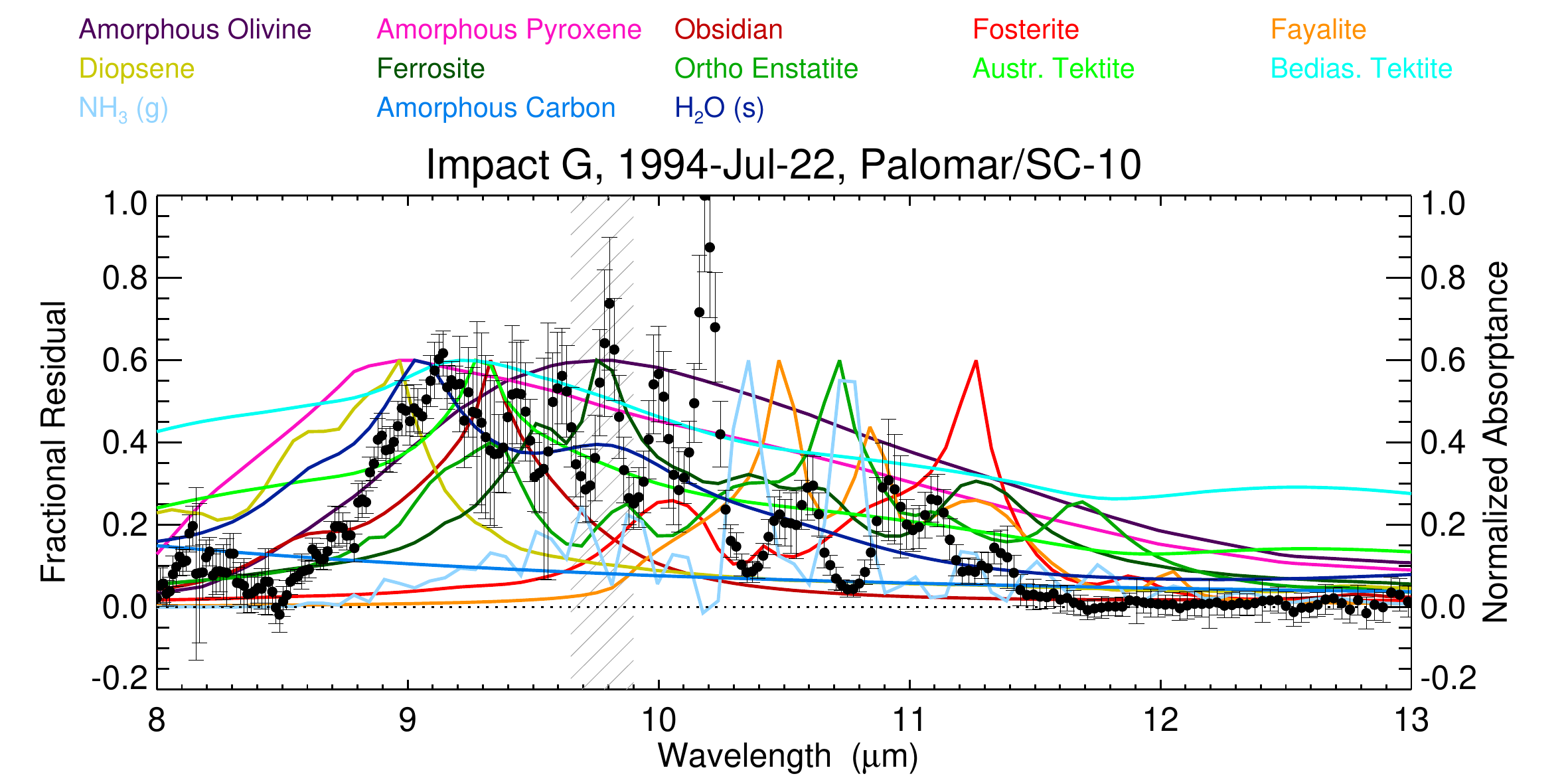}
\caption{Comparison of the fractional residual between observed and modelled SC-10 spectra of the SL9 G impact site, which denotes the non-gaseous component of the spectrum, and the absorptance spectra of candidate mineralogical species according to the legend.  The fractional residual was normalized to 1 and all mineral absorptance spectra have been normalized to 0.6 for ease of comparison. The 9.65 – 9.9 $\upmu$m range is hatched due to telluric ozone and so we do not consider any apparent features in this range. }
\label{fig:palomar_deammon}
\end{center}
\end{figure*}

Relative to the CIRS-derived \textit{a priori} temperature profile, heating of the atmosphere is evident from $\sim$100 mbar to $\sim$0.1 mbar (with no sensitivity to temperature at lower pressures).  The warmest temperatures of 174.3 $\pm$ 0.2 K were retrieved at the 2-mbar level.  The observations have limited sensitivity to temperature at pressures lower than 0.1 mbar and so we cannot rule out heating of the atmosphere at lower pressures.  At 30 mbar and 0.1 mbar, we derive NH$_3$ concentrations of $38^{+7}_{-19}$ ppmv and $5.7^{+4.5}_{-2.8}$ ppmv, respectively. The non-gaseous emission spectrum was derived by calculating the fractional residual (Equation \ref{eq:frac_resid}) between the observed spectrum and the best-fitting gaseous emission spectrum (Figure \ref{fig:palomar_example_fits} in blue).  Figure \ref{fig:palomar_deammon} compares the non-gaseous fractional residual with the features of candidate mineralogical species.  

\begin{figure*}[t!]
\begin{center}
\includegraphics[width=0.85\textwidth]{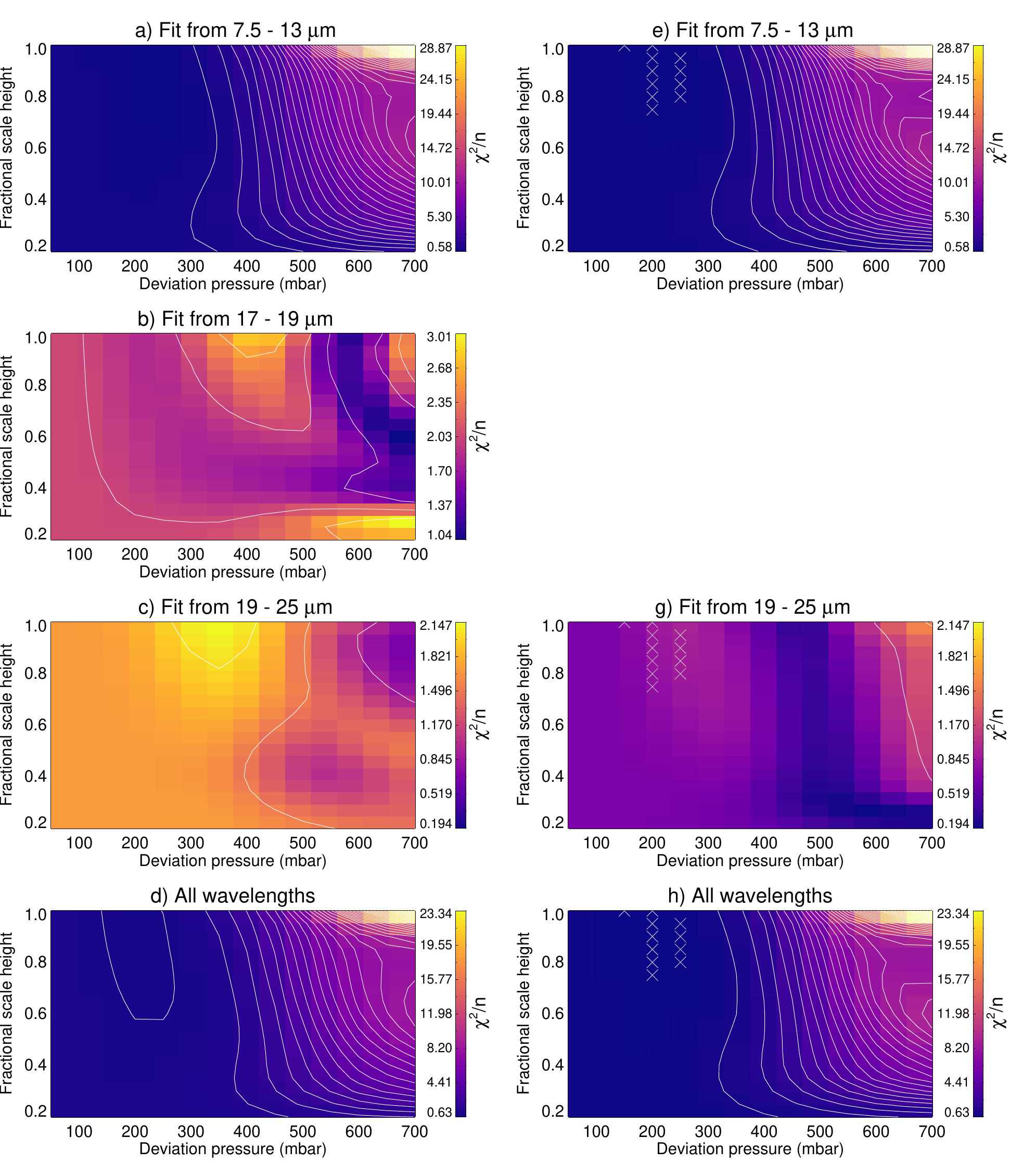}
\caption{Reduced $\chi^2$ distributions calculated from the observed and modelled T-ReCS spectra.  Results in the left column are for retrievals where the entire spectral range available (7.5 - 13, 17 - 25 $\upmu$m) was included in the inversion.  $\chi^2/n$ are shown at wavelengths from a) 7.5 – 13 $\upmu$m, b) 17 – 19 $\upmu$m, c) 19 – 25 $\upmu$m and d) all wavelengths.  Similar results are shown in panels e) – h) but where the 17 – 19 $\upmu$m range was excluded from the inversion. White contours denote increments of $\chi/n$ of 0.5.  White crosses denote model combinations that allow the 7.5 - 13 $\upmu$m and 19 - 25 $\upmu$m spectral ranges to be fit within (absolute) $\chi_{min}^2$ + 2.3, which corresponds to the 1-$\sigma$ confidence level.}
\label{fig:trecs_grid_results}
\end{center}
\end{figure*}

The derived non-gaseous emission exhibits a broad feature from wavelengths of 8.5 $\upmu$m to 11.5 $\upmu$m.  As in Section \ref{sec:mineral_modelling}, the dominant opacity sources responsible for the broad feature are a mixture of amorphous olivine, obsidian, and tekt2. Superimposed on this broad feature are narrower features at wavelengths of 9.6, 9.8, 10.0, 10.2, 10.6 and 10.9 $\upmu$m with (normalized) fractional residuals of 0.56 $\pm$ 0.12, 0.74 $\pm$ 0.16, 0.57 $\pm$ 0.12, 1.$\pm$ 0.18, 0.29 $\pm$ 0.08 and 0.30 $\pm$ 0.15.  We do not consider the feature at 9.8 $\upmu$m real due to its proximity to telluric O$_3$.  Although the observed feature at 10.0 $\upmu$m is consistent in wavelength with forsterite, the mineral exhibits a much broader feature (FWHM $\sim$0.4 $\upmu$m) compared to what is observed (FWHM $\sim$ 0.2 $\upmu$m).  In addition, forsterite has a feature at 11.25 $\upmu$m, which is twice as strong as its feature at 10.0 $\upmu$m, where no significant residual is observed.  In taking into account the 0.1 $\upmu$m (R = $\lambda/\Delta\lambda$ = 100) accuracy of the wavelength grid, the features at 10.6 and 11.0 $\upmu$m are near-coincident with those of fayalite at 10.5 and 10.9 $\upmu$m.  However, in acknowledging the 0.1-$\upmu$m accuracy of the wavelength grid, we also note that the features at 9.6, 9.8 10.0 and 10.2 $\upmu$m are within 0.1 $\upmu$m of NH$_3$ lines.  Thus, it remains inconclusive whether the identified narrow features result from non-gaseous species or poor fitting of the NH$_3$ emission features.

\subsubsection{Modeling Gemini/T-ReCS spectra}

The atmosphere derived from Cassini-CIRS measurements of 60$^\circ$S in 2001 (see \ref{sec:cirs_retrievals}) was adopted as the background atmosphere and \textit{a priori}.   The T-ReCS spectra were then inverted over the same model grid of NH$_3$ profiles adopted in inverting the SC-10 spectra (see Section \ref{sec:nemesis_sc10}).   In order to optimize the signal-to-noise ratio, we adopted the coadded spectrum in Figure \ref{fig:trecs_spx}, where all individual spectra sampling the impact region were coadded.  Initially, we inverted the entire T-ReCS spectral range from 7.5 to 13 $\upmu$m and 17 to 25 $\upmu$m.  Figure \ref{fig:trecs_grid_results}a-d shows the reduced $\chi^2$ distributions of the model grid in fitting the observations.

\begin{figure*}[t!]
\begin{center}
\includegraphics[width=0.8\textwidth]{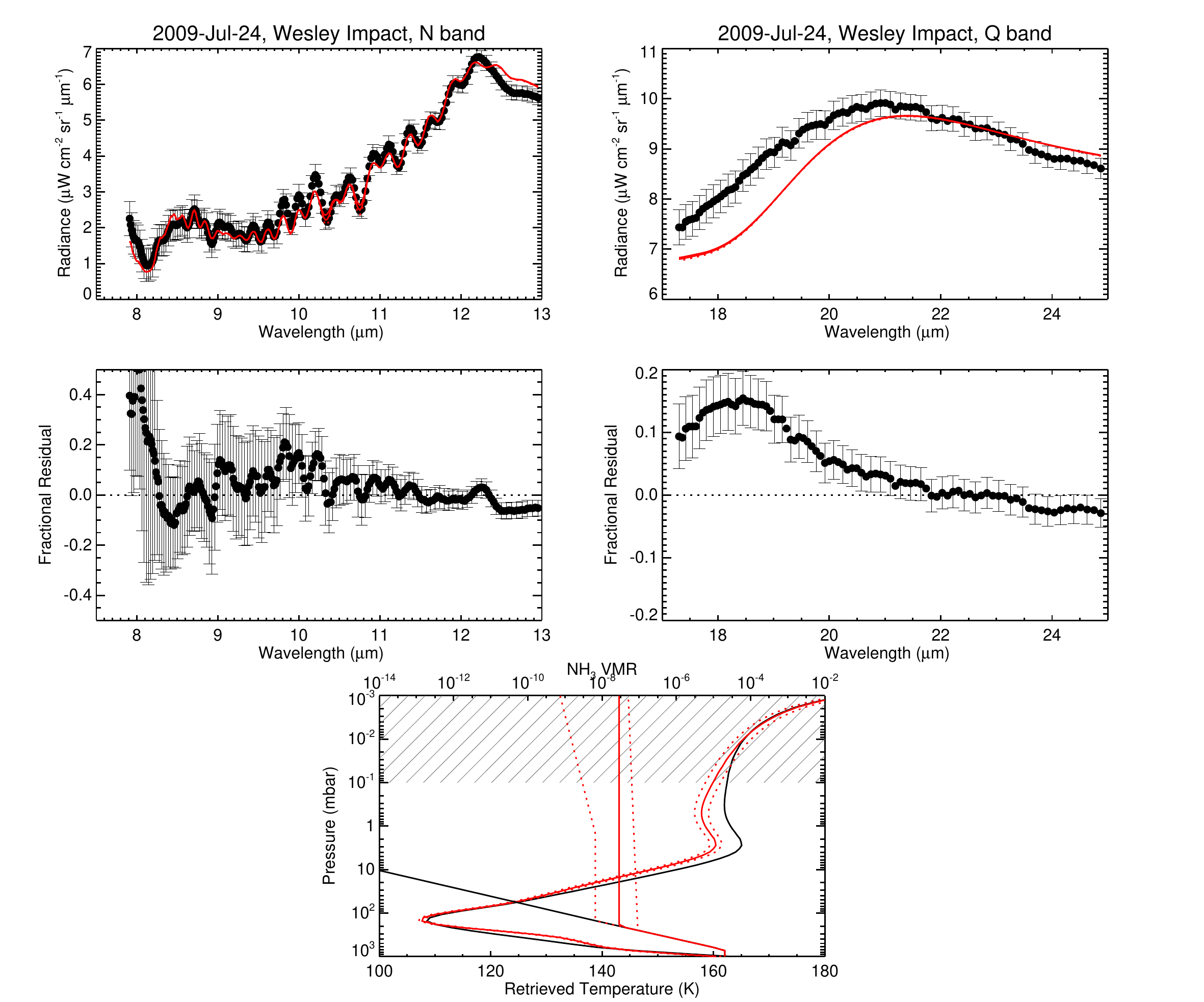}
\caption{ The top row compares observed T-ReCS spectra (black points with error bars) and the best-fitting synthetic spectra (solid, red lines) in the N (left) and Q (right) band.  The middle row shows the fractional residual between the observed and modelled spectra .  The bottom row shows the a priori (solid, back) and retrieved temperature profile (solid red, with 1-$\sigma$ uncertainty shown as dotted lines) according the lower axis.  The best-fitting NH$_3$ profile (solid black) is shown according the upper axis. Hashed regions at pressures lower than 0.1 mbar indicate where the observations have little/no sensitivity. }
\label{fig:trecs_example_fits}
\end{center}
\end{figure*}

Overall, the fit to the observations is optimized when the vertical profile of NH$_3$ deviates from the CIRS profile at 100 – 200 mbar and decreases in abundance with a high fractional scale height.   While we can adequately fit the 7.5 – 13 $\upmu$m spectrum with variations in temperature, NH$_3$ and tropospheric aerosol, we find that the 17 – 19 $\upmu$m feature cannot be adequately fit with a  physically-sensible atmospheric model and cannot be adequately fit using an atmospheric model that simultaneously fits the 7.5 – 13 $\upmu$m range.  This is further suggestive the 17 – 19 $\upmu$m enhancement of the Wesley impact site, with respect to the background atmosphere, results from a non-gaseous species in the atmosphere. 

As shown in Figure \ref{fig:trecs_grid_results}b-c, when the 17 – 19 $\upmu$m was included in the spectral inversion, we found that the overall fit to 17 – 25 $\upmu$m spectral range was poor.  We performed a second set of retrievals where the 17 – 19 $\upmu$m spectral range was omitted from the spectral inversion and found that the fit to the 19 – 25 $\upmu$m range significantly improved (comparing Figures \ref{fig:trecs_grid_results}c and g).  In the remainder of this section, we will therefore discuss results from retrievals omitting the 17 – 19 $\upmu$m range.

The best-fitting ($\chi^2/n$  = 0.629) NH$_3$ profile deviates from the CIRS-derived profile at $p_0$ = 200 mbar, and decreases with a fractional scale height of 1. However, we note that a range of solutions with p$_0$ ranging from 150 to 350 mbar and fractional scale heights from 0.75 to 1.0 also fit the observed spectrum with (absolute) $\chi^2$ values less than $\chi_{min}^2$ + 2.3, which corresponds to the 1-$\sigma$ confidence level when varying two parameters \citep{press_1992}.  These models are marked in Figure \ref{fig:trecs_grid_results}e – h. A fractional scale height of 1 imposes a constant abundance of NH$_3$ with altitude from the 200-mbar level to the top of the atmosphere, which is physically unrealistic.  Instead, the vertical profile of NH$_3$ is likely to be approximately constant with altitude from $p_0$ = 2 mbar to $\sim$ 0.1 mbar, over which the observations have sensitivity, but then decreases significantly at higher altitudes outside the vertical range of sensitivity of the observations.   We adopted the best-fitting spectrum (see above) as the gaseous component of the impact region.  The uncertainty of the gaseous spectrum was determined by calculating the the standard deviation of the synthetic spectra that had (absolute) $\chi^2$ less than $\chi_{min}^2$ + 2.3.  At each atmospheric level, the standard deviation in NH$_3$ values from the range of the model atmosphere that satisfied the aforementioned criteria were calculated and adopted as their respective uncertainties.  We performed a similar calculation to determine the temperature uncertainty but found that the standard deviation was smaller than the temperature uncertainty from the best-fitting retrieval and so the latter was adopted as the real temperature uncertainty.  

Figure \ref{fig:trecs_example_fits} compares the observed T-ReCS spectra with the gaseous spectrum noted above.  This result was forward modelled at 17 – 19 $\upmu$m (omitted from the inversion as noted above) so this spectral range could be included in the comparison.  At 30 mbar, we derive NH$_3$ abundances of  $150^{+338}_{-121}$ ppbv, which is in agreement (within uncertainty) of the results derived by \citet{fletcher_impact_2011}.   As also noted in previous work (e.g. \citealt{orton_2011}, \citealt{fletcher_impact_2011}), there is no evidence of stratospheric heating associated with the Wesley impact.  In fact, stratospheric temperatures are $\sim$10 K cooler than those derived from Cassini measurements in 2001.  In the Q band, the model spectra cannot adequately fit the observed spectra from 17.5 – 19.5 $\upmu$m.  For example, at 18.5 $\upmu$m, the fractional residual is 15.4 $\pm$ 4.3\% and therefore greater than 3$\sigma$. However, we find we can adequately fit the majority of the spectrum in the N band by varying only the vertical profiles of temperature, NH$_3$ and scaling the abundance of C$_2$H$_4$.

\section{Discussion}\label{sec:discuss}

Using methods as consistently as possible, our analysis of mid-infrared data capturing the Jovian atmosphere in the aftermath of the SL9 and Wesley impacts have revealed similarities and differences between the two events.  

In both impacts, NH$_3$, normally concentrated below Jupiter’s tropopause, was transported to stratospheric altitudes by ballistic splashback, which produced strong NH$_3$ emission features in the N band with the strongest lines at $\sim$10.4 and $\sim$10.8 $\upmu$m (see Figure \ref{fig:compare_all}, as seen in previous work \citep{griffith_1997,fast_2002,orton_2011,fletcher_impact_2011}.  In inverting the spectra of both impacts, we parameterized the NH$_3$ vertical profile by using the CIRS-derived profile (see \ref{sec:cirs_retrievals}) at pressures higher than a deviation pressure, $p_0$ and a decrease in abundance with altitude according to a fractional scale height, $f$.  The fit to the continua and the strength of the NH$_3$ emission lines in the Palomar/SC-10 spectra of SL9's G impact was optimized using $p_0$ = 650 mbar and a fractional scale height of 0.65.  This results in NH$_3$ concentrations of $38.4^{+6.8}_{-18.7}$ ppmv and $5.66^{+4.54}_{-2.82}$ ppmv at 30 and 0.1 mbar, respectively.  These are broadly-consistent with the findings of \citet{griffith_1997}, who derived $\sim$1 ppmv NH$_3$ concentrations at 0.1 mbar using data recorded $\sim$6 days after K impact, and \citet{fast_2002}, who derived $\sim$3 ppmv NH$_3$ concentrations at pressures lower than 1 mbar using data recorded $\sim$4 days after the G impact.  The fit to Gemini-South/T-ReCS spectra of the Wesley impact was optimized using $p_0$ = 200 mbar.  At 30 mbar, we derive NH$_3$ abundances of $150^{+338}_{-121}$ ppbv, which is in agreement within uncertainty of the results derived by \citet{fletcher_impact_2011}.   However, we note to the reader the our inversion of the spectra did not include the opacities of non-gaseous features and so these concentrations represent an upper limit.   Nevertheless,  the higher values of p$_0$ and higher concentrations of NH$_3$ required to fit the spectra of the SL9 impact, compared to the Wesley impact, is consistent with the SL9 fragments entering the atmosphere at angles closer to the local zenith, and reaching deeper, NH$_3$-richer altitudes.  

In both events, material introduced by the impactors into the atmosphere was evident as non-gaseous emission features in the spectra of the impact.  In order to determine the mineral species responsible for these non-gaseous emission features and their relative abundances, we performed a least-squares search of the mineral species listed in Table \ref{tab:mineral}.  In the impact sites of SL9, we would have expected a preponderance of amorphous olivine and pyroxene, plus water ice and amorphous carbon as well as NH$_3$ gas lofted to stratospheric altitudes.  The results of our mineralogical least-squares fitting indicate that amorphous olivine and NH$_3$ gas were significant opacity sources in the impact site of SL9's G fragment.  However, our analysis detected silica (in the form of obsidian) and an absence of amorphous pyroxene.  This was initially surprising since comets are expected to have similar abundances of olivinaceous to pyroxenacous species (e.g. \citealt{lisse_2006,lisse_2007b}).  We suggest that high temperatures and pressures produced by the cometary impacts converted all the cometary pyroxene into silicas.  While amorphous pyroxene is stable up to temperatures of $\sim$1000 K, amorphous olivine is stable up to $\sim$1500 K and so presumably, the material was heated to a temperature between 1000 - 1500 K such that amorphous olivine remained unaltered.  A similar finding was seen in the HD172555 exodisk system by \citet{lisse_2009}.  As far as we are aware, silica was not detected or reported in previous studies of the SL9 impact residue though this is unsurprising given the majority of mid-infrared studies focused on high-resolution spectroscopy of NH$_3$ lines between 10 - 11 $\upmu$m (e.g. \citealt{griffith_1997}, \citealt{fast_2002}) and did not sample the 9 - 9.5 $\upmu$m region.  \citet{nicholson_1995} presented Palomar/SC-10 spectra (the same dataset used in this paper) of the R impact, also observing a broad N-band feature, but this was interpreted as silicate emission, not silica.  However, their study did not attempt to derive an emissivity spectrum of the impact (by computing a residual between the impact and a non-impact region) or model the features using a least-squares search of mineral species. 

If the Wesley impactor was a comet, we would also have expected a composition of predominantly amorphous olivine and pyroxene, plus water ice and amorphous carbon.  If the Wesley impactor was an asteroid, as suggested previously (e.g. \citealt{hammel_2010, fletcher_impact_2010, orton_2011}, we would have expected a preponderance of crystalline olivine and phyllosilicates plus impact-alteration-produced silicas. Instead, our analysis found that the Wesley impact emission appears composed of almost pure amorphous olivine, with some ferromagnesian pyroxenes mixed in at an $\sim$8:1 ratio. No emission due to silicas (expected from 9.1 - 9.4 $\upmu$m) is apparent in our new reduction and spectral analysis, a marked difference from the findings of \citet{fletcher_impact_2011} who used the same dataset.  We believe our results differ due to differences in how the spectra were coadded and modelled.  \citet{fletcher_impact_2011} inverted spectra of the impact obtained over a range of emission angles and found the 9.1-$\upmu$m signature was most evident in the higher emission-angle observations, whereas, in this work, we averaged individual T-ReCS spectra over a range of emission angles to compute a single, coadded spectrum.   

Given the presence of silicas is analysis-dependent, we consider its non-detection in this work a tentative result.  If we assume the non-detection of silica is correct, we must therefore assume residue from the Wesley impactor did not reach the required temperatures of $\sim$1000-1500K in order to convert rocky silicates into silicas.   Instead, we must conclude from our new study that the Wesley residue was formed at low temperatures and high pressures such that the usually dominant crystalline olivine was readily converted into amorphous olivine. Whether these pressures were achieved due to the shock of impact, or due to the pressures achieved at depth inside a large differentiated body, is not clear. We currently favor the impact induced shock amorphization explanation, as all reports have the Wesley impactor as being relatively small in size ($<$10 km radius), which would be too small to thoroughly self-pressure alter its olivine throughout.  In addition, such an explanation would require removal of the body’s low pressure, near-surface crystalline olivine without heating the high pressure phase enough to produce silica. This still belabors the question of how a body impinging on Jupiter with $\sim$65 km/sec (Jupiter’s escape speed) relative velocity and a huge specific energy per kg of material can manage to NOT melt and transform its matter.  As noted above, SL9 was seen to create large amounts of glassy silica (a material not found in comets) from amorphous pyroxene. A remote possibility is that the Wesley 2009 impactor entered the atmosphere at an extremely shallow angle, which would allow material to spall off the impactor at relatively low temperatures. If we take the velocity of impact required to convert silicates into silicas as $\sim$10 km/sec (\citet{lisse_2009} and references therein), an impact angle of less than tan$^{-1}(10/65)$ $\sim$ 9$^\circ$ with respect to the local horizontal would be required in order to impart the impactors kinetic energy slow enough to the body that it could have a chance to retain all its silicates.  This is outside the 20$^\circ$ $\pm$ 5$^\circ$ estimated by \citet{sanchez_lavega_2010} and \citet{pond_2012a} in order to explain the lack of stratospheric heating \citep{fletcher_impact_2010} and the observed debris deposition between 3- 10 mbar \citep{dePater_2010,hammel_2010}.

An impact on Jupiter by 0.5- to 1.5-km objects, similar in size to the fragments of SL9 and the Wesley impactor, is likely in the near future.   While \citet{schenk_2004} inferred from the cratering record on the Galilean moons that SL9/Wesley-sized impacts on Jupiter will occur once every $\sim$100 years, \citet{sanchez_lavega_2010} instead calculated a rate of once every 10 - 20 years from the observed, recent rate.  For future impact events, we believe a mid-infrared characterization of the impact site would allow the impactor and its effect on the atmosphere to be characterized in the following ways.  First, the relative proportions of amorphous pyroxanaceous and olivinaceous species would be constrained from their features at 9, 10, and 19 $\upmu$m.  Second, the detection or non-detection of silica would indicate whether temperatures and pressures produced by the impact reached the $\sim$1500 K or 4 - 6 GPa levels required to temperature alter silicate species.  Third, if stratospheric NH$_3$ emission is detected, concentrations of NH$_3$ could be derived, which in turn would constrain the terminal depth of the impactor material.    While broadband imaging and low-resolution spectroscopy readily allowed non-gaseous features from the impactor to be identified, the coarse spectral resolving power (R = 100) presented in this work introduced uncertainty in identifying gaseous vs. non-gaseous spectral features and allowed only simple (two variable) parameterizations for the vertical profile of NH$_3$.  Characterization of future impacts would therefore benefit from N- and Q-band spectroscopy performed near-simultaneously and at higher spectral resolving powers (1000 $<$ 15000).  For Earth-based observations, R $>$ 1000 would also better allow CH$_4$ emissions from the planet to be disentangled relative to telluric CH$_4$ features (assuming the Earth-Jupiter relative velocity exceeds $\sim$ 15 km/s).  This would improve the retrieval of the stratospheric temperature field since the CH$_4$ is generally considered to be horizontally homogeneous in the lower stratosphere. 

\section{Conclusions}

We performed a retrospective analysis of mid-infrared Earth-based observations of Jupiter recorded in July 1994 and July 2009 in order to compare the effects on Jupiter’s atmosphere by Comet D/Shoemaker-Levy 9 (SL9) and the Wesley impactor quantitatively.  Spectrophotometry by IRTF/MIRAC and Palomar/Spectrocam-10 in 1994 and Gemini/T-ReCS in 2009 were reduced, calibrated and analyzed using methods as consistently as possible in order to facilitate robust comparisons between the two impacts.  As presented in previous work, we find the 8 – 11.5 $\upmu$m spectral range of both impact sites is significantly enhanced due to  stratospheric emission from NH$_3$, having been lofted from the upper troposphere to the stratosphere by ballistic ``blowout'', as well as non-gaseous impact material (see below).  Also, in agreement with previous work, we find that the SL9 impacts exhibited enhanced stratospheric CH$_4$ emissions whereas no significant enhancement was evident in the site above the Wesley impact.  In new findings, we determine that the sites of the SL9 and Wesley impacts both exhibit enhanced emissions at 18 – 19 $\upmu$m.  For example, at 17.9 $\upmu$m, the SL9 L impact site is enhanced by 41.2 $\pm$ 11.5\% and the Wesley impact site is by 16.4 $\pm$ 3.7\%. In order to detect and quantify the species responsible for these observed spectral features, we performed two separate but complementary analyses.  First, we performed mineralogical spectral modeling by adopting a collection of candidate mineralogical species and performing a grid search to find a combination of species that produced a model emissivity spectrum that minimized the goodness-of-fit to the observations. For the impact site of SL9's G fragment, we find that stratospheric NH$_3$ emission, amorphous olivine and silica (in the form of obsidian) are the dominant opacity sources responsible for the observed spectral features at 8.5 - 11 $\upmu$m and 18 - 19 $\upmu$m.  We find no evidence of amorphous pyroxene in the site of SL9's G fragment, which was initially unexpected since comets typically exhibit a $\sim$1:1 ratio of pyroxenaceous species to olivinaceous species.  We suggest that the high pressures and temperatures produced by the impact readily converted cometary pyroxene into silicas.  For the Wesley impact, we find that stratospheric NH$_3$ and amorphous olivine are the dominant opacity sources responsible for the observed spectral features.  We find no evidence of silicas in the aftermath of the Wesley impacts.   This is in contrast to previous work \citep{fletcher_impact_2011} who used the same data but processed/analyzed differently, which suggests the detection or non-detection of silica is analyssis-dependent. We therefore consider the non-detection of silica a tentative finding.  If the non-detection of silica is correct, it would require that material from the Wesley impactor was modified at relatively lower temperatures such that species retain all their silicates.  This would require that the Wesley impactor entered the atmosphere at a very shallow angle of 9$^\circ$, which is shallower than the 20 $\pm$ 5$^\circ$ estimated previously \citep{sanchez_lavega_2010,pond_2012a}.  Our second analysis involved inverting the spectra of both impacts using radiative transfer software in order to constrain NH$_3$ abundances.  For SL9's G impact, we constrain NH$_3$ concentrations of 38.4$^{+6.8}_{-18.7}$ and 5.66$^{+4.54}_{-2.82}$ ppmv at 30 and 0.1 mbar, respectively, which is broadly consistent with previous work \citep{griffith_1997,fast_2002} though our parameterization of NH$_3$'s vertical profile differed.  For the Wesley impact, we derive an NH$_3$ concentration of 150 $^{+338}_{-121}$ ppbv at 30 mbar, which is consistent (within uncertainty) with \citet{fletcher_impact_2011}.  Higher concentrations of NH$_3$ in the SL9 impacts is consistent with the fragments reaching a deeper terminal depth compared to the Wesley impact. 

\section{Acknowledgements}

Some of this research was carried out at the Jet Propulsion Laboratory, California Institute of Technology, under a contract with the National Aeronautics and Space Administration (80NM0018D0004). During the course of this research, Meera Krishnamorthy was a student at the California Institute of Technology and worked at JPL as an intern in the Caltech Summer Undergraduate Research Fellowship (SURF) program, supported by these funds.  Fletcher was supported by a European Research Council Consolidator Grant (under the European Union's Horizon 2020 research and innovation programme, grant agreement No 723890) at the University of Leicester.  We thank Mike Chaffin and Sonal Jain for creating IDL-compatible color tables\footnote{https://github.com/planetarymike/IDL-Colorbars} that are perceptually-uniform, colorblind-friendly, and print correctly in black and white.  Emma Dahl was supported by an appointment to the NASA Postdoctoral Program at the Jet Propulsion Laboratory, administered by Oak Ridge Associated Universities under contract with NASA.

\section{Data Availability}

Archiving of the IRTF/MIRAC images and the Gemini-South/T-ReCS spectra at the Planetary Data System (PDS) is in progress (at the time of writing).  Until they are available on the PDS, the data will be made available upon request.  The calibrated spatial-spectral Palomar/SC-10 image of Jupiter is available at the Mendeley data archive\footnote{DOI:10.17632/ydt2ff9n83.1}.  Raw data files and the software used to process the Palomar data would also be made available upon request.

%\bibliography{ref}

\appendix

\section{Cassini-CIRS retrievals}\label{sec:cirs_retrievals}

During Cassini's 2000/2001 flyby of Jupiter, four maps of 2.5 cm$^{-1}$ `MIRMAP' observations - `ATMOS02A', `ATMOS02B', `ATMOS02C' and `ATMOS02D' - were acquired from January 1st to Jan 11th.  Spectral from all four maps were sorted into 2$^\circ$ latitude bins, nyquist-sampled by 1$^\circ$S and coadded.  The effective noise on these spatially-binned observations was assumed to be the largest of either: 1) the standard deviation of the mean or 2) the noise-equivalent spectral radiance (NESR) spectrum of CIRS by a factor of $(0.5/{\Delta\tilde{\nu}}) (1/(\sqrt{N}) + 1/(\sqrt{M}))$, where $\Delta{\tilde{\nu}}$ is the spectral resolution of the target spectra, N is the number of averaged target spectra and M is the number of deep-space spectra.  

The spectra at 40$^\circ$S (the latitude of the SL9 impacts) and 60$^\circ$S (the latitude of the Wesley impact) were inverted by varying temperature, NH$_3$, tropospheric aerosol, C$_2$H$_2$, C$_2$H$_4$ and C$_2$H$_6$.  Inversions were performed using the NEMESIS radiative transfer code \citep{irwin_2008}.  The sources of line data, the atmospheric model are identical to those described in Section \ref{sec:inversions}.  

The vertical temperature profile was allowed to vary continuously at all altitudes.  The vertical profile of NH$_3$ was parameterized with three variables: 1) a deep volume mixing ratio ($q_{deep}$) at pressures higher than a knee pressure, 2) the knee pressure ($p_{knee}$) and 3) a fractional scale height ($f_{NH3}$).  For tropospheric aerosol, a grey absorber was assumed with an \textit{a priori} base pressure ($p_{base}$) of 0.7 bar, an optical depth ($\tau_1$) of 1 and a fractional scale height ($f_{cloud}$) of 0.2, were assumed, and all three parameters were varied. The vertical profiles of C$_2$H$_2$, C$_2$H$_4$ and C$_2$H$_6$ were adopted from a photochemical model of Jupiter \citep{moses_2017} and varied by applying a scale factor to all altitudes.

\begin{figure*}[t!]
\begin{center}
\includegraphics[width=0.8\textwidth]{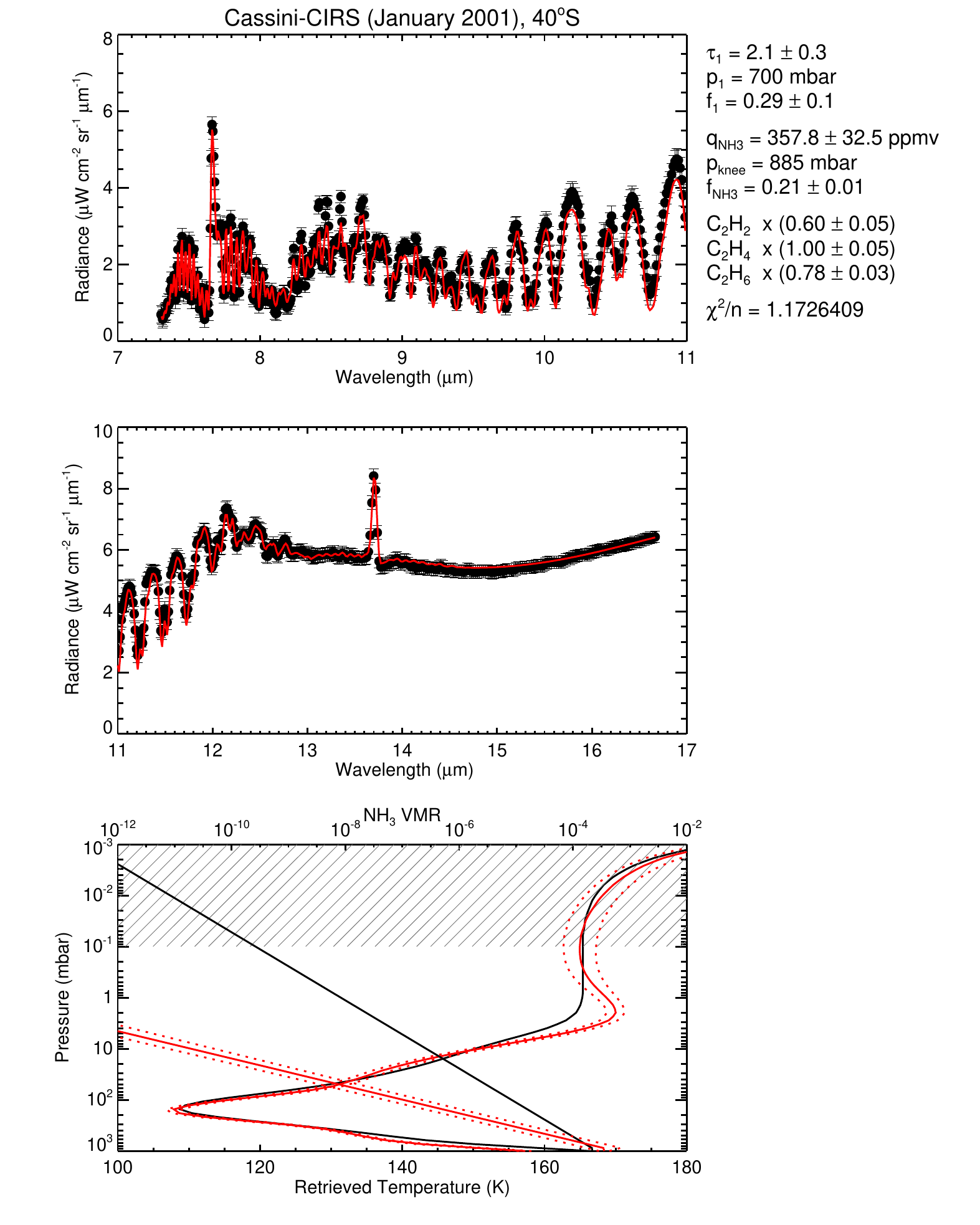}
\caption{Inversions of Cassini-CIRS spectra recorded during the 2001 Jupiter flyby at 40$^\circ$S.  The top and middle panels show observed spectra (black circles with error bars) from 7 - 11 $\upmu$m and 11 - 17 $\upmu$m, and synthetic spectra are shown as solid lines.  The retrieved atmospheric parameters corresponding to the synthetic spectra are shown in the bottom panel and the legend.  Black, solid lines represent \textit{a priori} values, solid lines denote retrieved values and dotted lines outline the 1-$\sigma$ uncertainty.}
\label{fig:plot_40S_mirmap}
\end{center}
\end{figure*}

\begin{figure*}[t!]
\begin{center}
\includegraphics[width=0.8\textwidth]{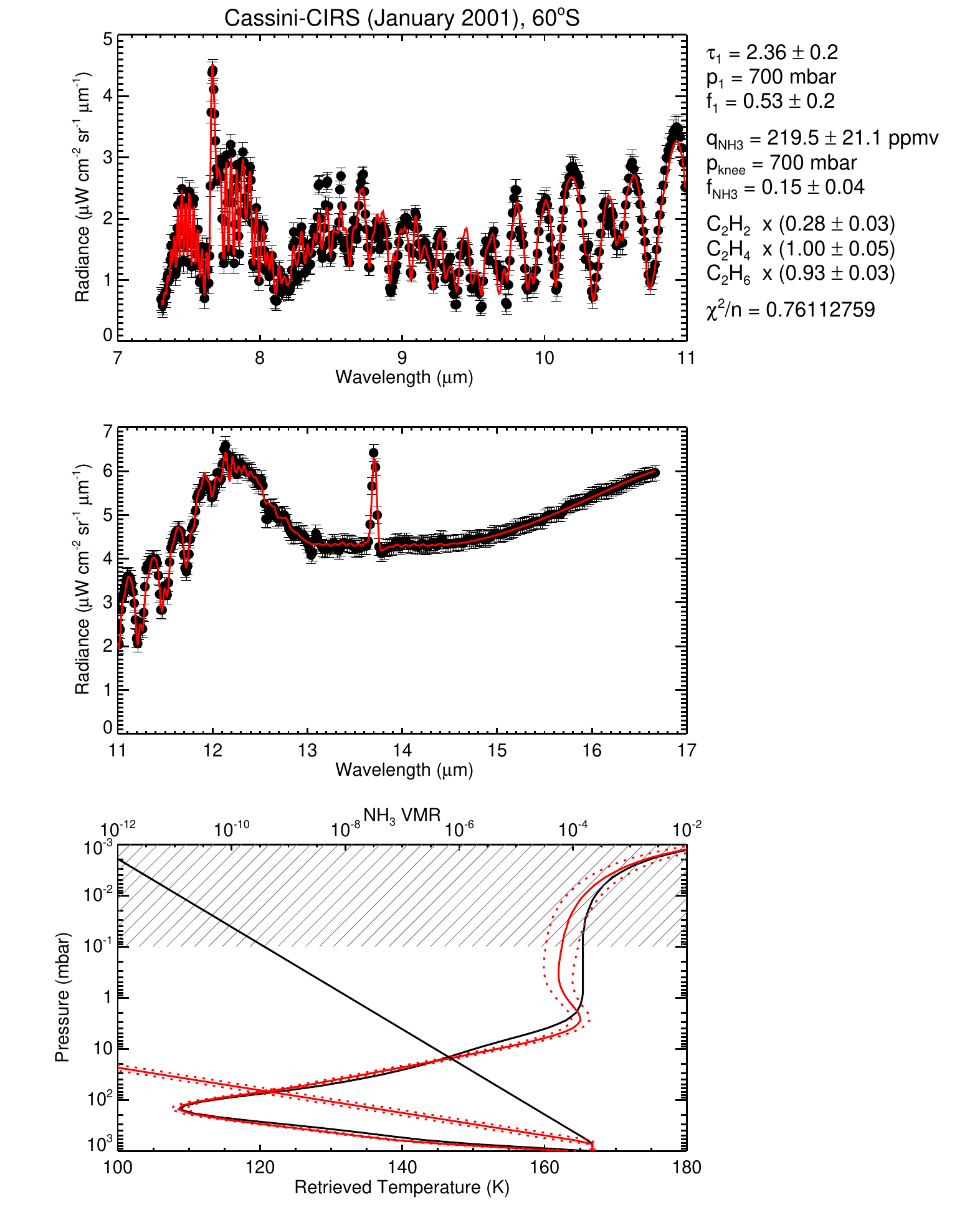}
\caption{Inversions of Cassini-CIRS spectra recorded during the 2001 Jupiter flyby at 60$^\circ$S.  The top and middle panels show observed spectra (black circles with error bars) from 7 - 11 $\upmu$m and 11 - 17 $\upmu$m, and synthetic spectra are shown as solid lines.  The retrieved atmospheric parameters corresponding to the synthetic spectra are shown in the bottom panel and the legend.  Black, solid lines represent \textit{a priori} values, solid lines denote retrieved values and dotted lines outline the 1-$\sigma$ uncertainty.}
\label{fig:plot_60S_mirmap}
\end{center}
\end{figure*}

Comparisons of the observed and synthetic spectra and the corresponding retrieval parameters are shown in Figures \ref{fig:plot_40S_mirmap} and \ref{fig:plot_60S_mirmap}.  The retrieved atmospheric parameters at 40$^\circ$S and 60$^\circ$S are adopted as \textit{a priori} values for inversions of the Palomar/SC-10 and Gemini-S/T-ReCS spectra, respectively.

\end{document}